\documentclass{aa}  
\usepackage[dvipsnames]{xcolor}
\usepackage{layouts}
\usepackage{graphicx}
\usepackage{todonotes}
\usepackage{rotating} 
\usepackage{caption}
\usepackage{placeins}
\usepackage{rotating}

\usepackage{xspace}
\def\aarm{TDE~2025aarm\xspace}

\usepackage[breaklinks=true]{hyperref}
\usepackage{natbib,twoopt}
\bibpunct{(}{)}{;}{a}{}{,} 

\bibpunct{(}{)}{;}{a}{}{,}             
\makeatletter
  \newcommandtwoopt{\citeads}[3][][]{\href{http://adsabs.harvard.edu/abs/#3}%
    {\def\hyper@linkstart##1##2{}%
     \let\hyper@linkend\@empty\citealp[#1][#2]{#3}}}
  \newcommandtwoopt{\citepads}[3][][]{\href{http://adsabs.harvard.edu/abs/#3}%
    {\def\hyper@linkstart##1##2{}%
     \let\hyper@linkend\@empty\citep[#1][#2]{#3}}}
  \newcommandtwoopt{\citetads}[3][][]{\href{http://adsabs.harvard.edu/abs/#3}%
    {\def\hyper@linkstart##1##2{}%
     \let\hyper@linkend\@empty\citet[#1][#2]{#3}}}
  \newcommandtwoopt{\citeyearads}[3][][]%
    {\href{http://adsabs.harvard.edu/abs/#3}
    {\def\hyper@linkstart##1##2{}%
     \let\hyper@linkend\@empty\citeyear[#1][#2]{#3}}}
\makeatother
\usepackage{graphicx}
\usepackage{booktabs}
\usepackage{lipsum}
\usepackage{txfonts}
\usepackage{hyperref}
\hypersetup{
    colorlinks=true,
    allcolors=Blue
}
\usepackage{tabularx}
\usepackage{upgreek}

\graphicspath{{Figures/}} 

\begin{document} 
\abstract
{We report the X-ray and optical spectroscopic properties of \aarm, the second closest tidal disruption event (TDE) discovered to date. The proximity of this source (60~Mpc), combined with a deep and intense X-ray-monitoring campaign spanning six months, allowed us to probe the source down to an unprecedented 0.2--10~keV luminosity of $\sim7\times10^{39}$ erg s$^{-1}$ at times close to the optical peak. This renders \aarm the faintest early-X-ray-detected TDE to date. After the first X-ray detection, the source brightened by nearly two orders of magnitude, reaching a peak luminosity of $\sim5\times10^{41}$ erg s$^{-1}$ about four months after the optical peak. Through time-resolved X-ray spectral analysis, we find that \aarm evolved from an initially hard, power-law-dominated X-ray state into a softer, disk-dominated state as the luminosity increased, before hardening again at later times. Such low-hard-to-high-soft state transitions are commonly observed in black hole X-ray binaries (XRBs) but have not been reported in thermal TDEs  previously. We show that the spectral evolution can be described by variations in the relative contributions of an accretion disk and a Comptonizing component, qualitatively resembling the disk--corona evolution observed in XRBs. We also present the results of our optical spectroscopic follow-up campaign with HET/LRS2, confirming the TDE classification and revealing N\,III Bowen fluorescence features. The extremely faint early-time X-ray emission of \aarm further supports the idea that the historical dichotomy between X-ray-bright and X-ray-undetected TDEs is largely driven by selection effects related to the depth, cadence, and duration of X-ray follow-up observations. \aarm therefore provides new insight into both the accretion physics of TDEs and the possible universality of accretion across several orders of magnitude in black hole mass.}



\authorrunning{Pietro Baldini et al.}
\title{Low-hard to high-soft spectral state transitions in the faintest early X-ray-detected optical tidal disruption event TDE 2025aarm}

\titlerunning{The X-ray state transitions of \aarm.}
\author{P. Baldini\inst{1,}\thanks{corresponding author, \href{mailto:baldini@mpe.mpg.de}{baldini@mpe.mpg.de},}
  \and A. Rau\inst{1}
  \and A. Merloni\inst{1}
  \and J. Somalwar\inst{2,3}
  \and S. J. Brennan\inst{1}
  \and E. Kyritsis\inst{1}
  \and H.~C.~I.~ Wichern\inst{4}
  \and P. Boorman\inst{1}
  \and P. Charalampopoulos\inst{5,6}
  \and L. Dai\inst{7,8}
  \and G. K.  Jaisawal\inst{4}
  \and C. Jin\inst{9}
  \and T. Lian\inst{1}
  \and K. Nandra\inst{1}
  } 

\institute{Max-Planck-Institut f\"ur extraterrestrische Physik, Giessenbachstra{\ss}e 1, D-85748 Garching bei M\"unchen, Germany
\and Department of Astronomy, University of California, Berkeley, CA 94720, USA
\and Kavli Institute for Particle Astrophysics \& Cosmology, P.O. Box 2450, Stanford University, Stanford, CA 94305, USA
\and DTU Space, Department of Space Research and Space Technology, Technical University of Denmark, Elektrovej 327, 2800 Kgs.
Lyngby, Denmark
\and Institute of Space Sciences (ICE-CSIC), Campus UAB, Carrer de Can Magrans, s/n, E-08193 Barcelona, Spain
\and Finnish Centre for Astronomy with ESO (FINCA), FI-20014 University of Turku, Finland
\and Department of Physics, The University of Hong Kong, Pokfulam Road, Hong Kong, China
\and The Hong Kong Institute for Astronomy and Astrophysics, The University of Hong Kong, Pokfulam Road, Hong Kong, China
\and National Astronomical Observatories, Chinese Academy of Sciences, Beijing 100101, China}

\maketitle

\section{Introduction}

Tidal disruption events (TDEs) are the result of the unlucky encounter between a star and a supermassive black hole (SMBH). When the tidal forces overcome the star’s self-gravity, the star is torn apart, and its debris is accreted onto the black hole (\citealp{hills1975possible,rees1988tidal,evans1989, Phinney1989}, and see \citealp{gezari2021tidal} and \citealp{jonker2021tidal} for extensive reviews). This process produces transient X-ray-through-radio emission flaring on timescales of a few weeks and lasting for several months. TDEs are therefore unique probes of dormant SMBHs, complementing the SMBH population revealed by active galactic nuclei (AGN). Moreover, they are the perfect environment for studying the formation, evolution, and exhaustion of accretion flows onto SMBHs.  

The first TDEs were discovered in the 1990s with ROSAT (\citealp{trumper1982rosat, komossa1999discovery, komossa1999giant, grupe1999rx, greiner2000rx}) as transient soft X-ray flares. While a few more candidates were then discovered through other X-ray telescopes, such as Chandra, the X-ray Multi-Mirror Mission (XMM-Newton), and Swift (\citealp{weisskopf2000chandra, jansen2001xmm, Gehrels2004}, and see \citealp{komossa2015tidal, Saxton2020}), over the past two decades, the vast majority of candidates have been discovered in the optical/UV thanks to time-domain optical surveys such as the Panoramic
Survey Telescope and Rapid Response System (Pan-STARRS), tha Palomar Transient Facility (PTF), the All-Sky Automated Survey for Supernovae (ASAS-SN), the Asteroid Terrestrial-impact Last Alert System
(ATLAS), and the Zwicki Transient Facility (ZTF, \citealp{Chambers2016, Law2009, Kulkarni2013, shappee2014man, kochanek2017all,tonry2018atlas, bellm2018zwicky}). In the last few years, however, the advent of high-grasp time domain X-ray facilities, namely the extended ROentgen Survey with
an Imaging Telescope Array (eROSITA; \citealp{predehl2021erosita}), aboard the Spektrum–Roentgen–Gamma (SRG; \citealp{sunyaev2021srg}) satellite, and more recently the Einstein Probe mission (EP, \citealp{ep2022}), have started filling the gap between optically selected and X-ray-selected TDEs (e.g., \citealp{sazonov2021first,khorunzhev2022search, Grotova2025a, Grotova2025b, jin2025}). Currently, we know of about $200$ TDEs, of which one quarter were selected in the X-rays.

 Ensuring that selection is not biased toward one specific wavelength range is crucial, as the interplay between optical/UV and X-ray radiation is a powerful test of emission models, and given that TDEs are often detected to be flaring in either only X-ray or only optical/UV (\citealp{Guolo2024systematic, Grotova2025b}). In fact, while the X-ray emission is generally understood to originate from the inner regions of the newly formed disk, the optical/UV emission appears to originate from a larger and colder region with respect to what one would expect from accretion processes (e.g., \citealp{gezari2021tidal}). Because of this, several models have been proposed to explain the origin of the optical/UV emission in TDEs, which can be broadly lumped into two categories: 1) models in which the primary X-ray radiation is reprocessed to optical wavelengths by an optically thick envelope or outflow (e.g., \citealp{Guillochon2014, roth2016x, metzger2016bright, dai2018, parkinson2022optical}); and 2) models in which the optical emission is not powered by accretion, but it is rather produced by shocks within the debris stream  (e.g., \citealp{shiokawa2015general, jiang2016A, ryu2023shocks, steinberg2024,huang2025x, meza2025, Martire2025}). While no theory is currently able to reproduce all observables, it is increasingly clear that reprocessing alone is insufficient, at least in the first phases of the emission (e.g., \citealp{mummery2023x, mummery2025tidal}).

Historically, it was believed that two distinct populations of optically selected TDEs existed: X-ray-bright and X-ray-faint TDEs. This is because $\gtrsim50\%$ of optically selected TDEs are not detected in the X-rays (\citealp{Guolo2024systematic}). However, the systematic analysis by \citet{Guolo2024systematic} of the X-ray properties of optically selected TDEs has revealed that this is likely due to a selection bias. As a matter of fact, the optical-to-X-ray luminosity ratio of TDEs $L_{BB}/L_X$ spans several orders of magnitude continuously, exploring values of $L_{BB}/L_X = [0.5, 3000]$, in contrast with AGN, for which a tight correlation is found (known as $\alpha_{OX}$, e.g., \citealp{lusso2016}). Additionally, the X-ray peak for TDEs can be reached several months after the optical peak. The large population of X-ray undetected optical TDEs is therefore the result of shallow, and unsystematically cadenced X-ray follow-up of the more extreme end of the $L_{BB}/L_X$ distribution. Deep X-ray observations of TDEs are crucial for obtaining a complete picture of their emission mechanisms. 

Tangentially, over the past 10 years, such deep X-ray observations of TDEs have revealed a sub population showing the development at late times of a non thermal spectral component in addition to the accretion disk (e.g., \citealp{Wevers2021,yao2022, liu2024rapid, cao2024, chatterjee2026temporal, berger2026}, and see \citealp{chakramaster026} for a sample study). This component, which is well modeled by a power law, has been interpreted as a hot population of electrons, known as the corona, upscattering a fraction of the seed disk photons through inverse-Compton processes. This component is at the origin of the hard X-ray emission of AGN and of X-ray binaries (XRBs, e.g., \citealp{done2007modelling, gilfanov2014observational}). The emergence of a corona in TDEs is particularly valuable, as it allows us to probe the formation of such a poorly understood component in the case of SMBHs, since AGN generally already have a corona in place (but see e.g., \citealp{ricci2020}). Moreover, there have been attempts to link the development of the corona in TDEs to the so-called state transitions observed in the outbursts of black hole XRBs (e.g., \citealp{Wevers2021, Sfaradi2022, Sfaradi2025, berger2026}). In the case of black hole XRBs, time-resolved X-ray spectroscopy reveals that state transitions are due to a variable relative contribution of disk and corona, linked to the accretion rate evolution (see \citealp{belloni2010} for a review). Establishing a robust analogy between TDE and XRB state transitions would constitute compelling evidence for the scale invariance of accretion across several orders of magnitude in black hole mass, which has long been sought after (e.g., \citealp{merloni2003fundamental, mchardy2006active, noda2018explaining, ruan2019}). Thus far, however, the connection is only partial, with sources showing only the terminal high-soft-to-low-hard (or disk-dominated-to-corona-dominated) state transition. No TDE to date has been reported to show the characteristic initial XRB-like low-hard-to-high-soft (or corona-dominated-to-disk-dominated) state transition.

In this paper, we present the X-ray and optical properties of the TDE \aarm (first presented in \citealp{simongini2026early}). The unique proximity of this event has enabled unprecedentedly deep X-ray follow-up of this source, which is the faintest X-ray-detected TDE to date. In addition, as reported in this work, \aarm is the first known case of a TDE exhibiting low-hard-to-high-soft X-ray state transitions. In section \ref{Disc} we discuss the discovery and follow-up of \aarm, while in Sections \ref{sec:xray} and \ref{sec:opt} we discuss respectively the analysis of the X-ray and optical data. Lastly, we discuss our results in Section \ref{discussion}. Unless otherwise stated, all uncertainties in this work correspond to 68\% (1$\sigma$) confidence intervals.

\section{Discovery and follow-up}
\label{Disc}

\begin{figure*}
\centering
  \includegraphics[width=\linewidth]{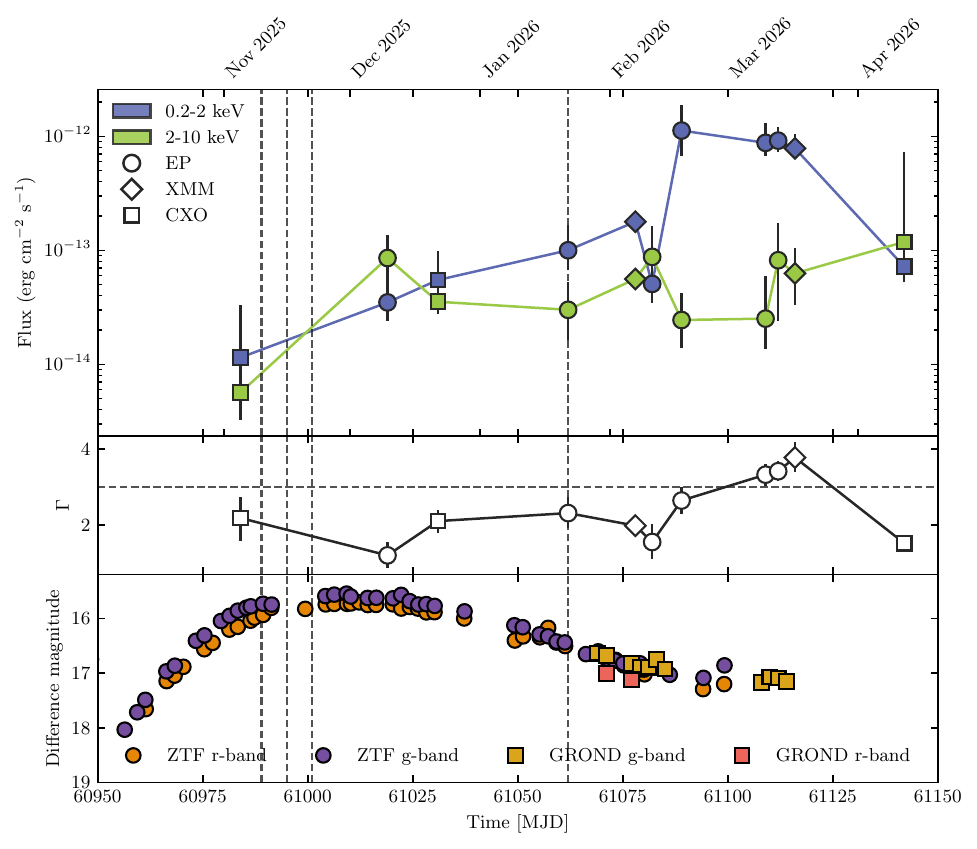}
  \caption{X-ray and optical light curve of \aarm. In the top panel, we plot the X-ray light curve in the 0.2-2 and 2-10 keV bands. In the middle panel, we show the evolution of the photon index $\Gamma$ over time (see text for more details about the spectral analysis). While the flux is derived from the more complete \textsc{diskbb+powerlaw} model, $\Gamma$ is derived from the single powerlaw fit, as a neutral indicator spectral shape (see Sect. \ref{subsec:xrayspec}). We indicate with the dashed line a photon index value of $\Gamma=3$, as this is the canonical photon index threshold to distinguish between thermal and non thermal emission in accreting SMBHs (e.g., \citealp{Grotova2025a, Grotova2025b}). The bottom panel shows the optical light curve collected with ZTF and GROND. The vertical dashed lines correspond to the epochs at which optical spectra were collected.
}
  \label{fig:lc}
\end{figure*}

As reported in \citet{simongini2026early}, \aarm was discovered on October 1st 2025 by the Gravitational-wave OpticalTransient Observer (GOTO, \citealp{steeghs2022gravitational}), as an optical transient in the nucleus of the galaxy LEDA 3681212 at z=0.0135 \citep{faris2025}. As the transient kept on rising over the following weeks, it was classified as a TDE because of its optical spectrum characterized by a blue continuum with strong and broad H+He lines \citep{newsome2025}. \citet{simongini2026early}, however, points out that possible N\,III\,$\lambda$4100 emission is present, which tentatively renders \aarm a member of the N\,III Bowen TDEs (e.g., \citealp{charalampopoulos2022}). We confirm such a classification in section \ref{sec:opt}.

We began monitoring \aarm with Einstein Probe's Follow-up X-ray Telescope (FXT, \citealp{ep2022}) on November 3rd 2025 until March 13th 2026, when the source went into sunblock (the X-ray observation log can be seen in Table \ref{tab:obslog}, and the light curve is presented in Fig. \ref{fig:lc}).
While our first observations did not yield any detection, a point-source spatially coincident with the location of the optical transient was discovered, as the result of a 25ks Chandra observation performed on November 5th 2025 (\citealp{somalwar}). The source was detected with a reported luminosity of 2.5$\times10^{39}$\, erg/s in the 0.5-7 keV band. None of the December EP observations detected emission from \aarm individually. However, stacking the December 7th and December 13th observation, resulted in a detection of an X-ray point source at the location of \aarm. Two more Chandra DDTs were collected on December 22nd and April 11th, the latter of which was reported in \citet{jaisawal}. We also obtained two XMM-Newton DDT observations, motivated by the unusual X-ray spectral behavior of \aarm on February 7th, 2026, and March 17th, 2026. We do not include any of the Swift/XRT observations, which were obtained by several PIs throughout the evolution of \aarm, as they do not meaningfully inform the X-ray evolution of the TDE, even when observations are stacked \citep{simongini2026early}. The Swift/XRT observations, however, are discussed together with the first two Chandra observations in \citealp{matsumoto2026faint}. We return to the results of this work in Section \ref{discussion}

In addition to the X-ray monitoring, we also collected four optical spectra of \aarm using the Low-Resolution Spectrograph 2 (LRS-2, \citealp{chonis2016lrs2}) on the Hobby–Eberly Telescope (HET, \citealp{ramsey1988het}), located on the McDonald Observatory in Davis Mountains, Texas, USA, probing both the rising and decaying phase of \aarm. 
The spectra were collected on November 10th, 16th, and 22nd, 2025, and on January 19th, 2026, thanks to the HET scheduling system \citep{shetrone2007ten}. The spectral reduction and analysis are presented and discussed in Section \ref{sec:opt} and Appendix \ref{app:opt}. Finally, \aarm was observed with the Gamma-Ray burst Optical Near-infrared Detector (GROND; \citealt{Greiner:2008}) mounted at the MPG 2.2\,m telescope at ESO’s La Silla observatory in twelve epochs between 2026 January 17 and April 14th. We discuss the observations and data reduction in Appendix \ref{app:opt}, and present the photometric points in Fig. \ref{fig:lc}.

\begin{table}[]
\caption{X-ray observation log of \aarm.}
\begin{tabular}{lllll}
\toprule
Date       & Telescope   & ObsID                                                                     & Exp. & ID \\ \midrule
25-11-03 & EP/FXT      & 06800000962                                                               &     2.99  &      \\
25-11-05 & Chandra & 31982                                                                         &     24.75 & CXO1 \\ 
25-11-07 & EP/FXT      & 06800000962                                                               &     0.4          \\

25-12-07 & EP/FXT      & 11900505216                                                               &     3.59  & EP1 \\
25-12-13 & EP/FXT      & 11900512512                                                               &     3.8   & EP1 \\
25-12-18 & EP/FXT      & 11900517120                                                               &     0.54        \\

25-12-22 & Chandra & 32075                                                                         &     15.03 & CXO2        \\

26-01-22 & EP/FXT      & 06800001129                                                               &     5.92 & EP2        \\
26-02-07 & XMM     & 0974790901                                                               &     24.0 & XMM1     \\

26-02-11 & EP/FXT      & 11900599424                                                               &     1.21 & EP3       \\
26-02-18 & EP/FXT      & 11900610048                                                               &     3.00 & EP4      \\

26-03-10 & EP/FXT      & 08500000576                                                               &     5.41  & EP5     \\
26-03-13 & EP/FXT      & 11900599424                                                               &     4.89  & EP6    \\

26-03-17 & XMM      & 0974791701                                                                &     14.25 & XMM2    \\

26-04-11 & Chandra& 32275                                                                    &     14.74  & CXO3       \\ 
 \midrule
\end{tabular}
\tablefoot{ The exposure time "Exp." is expressed in ks. The ID column shows how we refer to the individual observations in the text. Observations without an ID were reported for completeness but were not used in the analysis, as they resulted in non-detections for which the upper limits lacked constraining power. Two EP/FXT observations are labeled EP1, as we stacked them for our analysis.}

\label{tab:obslog}
\end{table}

\section{X-ray analysis}
 \label{sec:xray}

We reprocessed all X-ray event files using standard prescriptions. For Einstein Probe, we use the FXT Data Analysis Software Package (FXTDAS) v1.30, for Chandra we use CIAO v. 4.16.0, and for XMM-Newton we use SAS v. 21.0. We extracted source and background spectra from the processed event files using \textsc{xselect} tool of the \textsc{heasoft} software package, SAS v. 21.0, and CIAO v. 4.16.0, respectively, for EP/FXT, XMM-Newton/EPIC, and Chandra/ACIS-S. We used source extraction regions of 20" radius for EP/FXT, and XMM-EPIC pn, MOS1, and MOS2 instruments, while we used a radius of 1.5" for Chandra ACIS, centered on the location of the nuclear X-ray point source. We use nearby circular regions to extract background spectra.

All spectra were analyzed with the Bayesian X-ray Analysis software (BXA) version 4.1.2 (\citealp{buchner2014bxa}), which connects the nested sampling algorithm UltraNest (\citealp{buchner2019ultranest,buchner2021ultranest}) with the fitting environment CIAO/Sherpa (\citealp{fruscione2006sherpa}). Spectra were fit unbinned, using C-statistic. The fitting procedure included a PCA-based background model derived from a large sample of background spectra for each of the analyzed instruments (\citealp{simmonds2018}). As the EP/FXT hardware is analogous to that of the telescope modules of SRG/eROSITA, in the absence of a PCA-based background model for FXT we used that derived from a large set of eROSITA spectra (\citealp{liu2022efed}). While EP and eROSITA are on very different orbits, a visual inspection of the fits confirms that this approximation does not affect our results (see Fig. \ref{fig:xrayspecall}). For the EP/FXT spectra, we simultaneously fitted the FXT-A and FXT-B modules by including a normalization constant in all models. 

The first two EP observations (November 3rd and 7th) yielded no detection. As at a similar date, Chandra detected emission from the nucleus of the host galaxy down to fluxes unprobed by EP/FXT, we do not further discuss these data. None of the three EP observations collected in December significantly detected a source at the location of \aarm; however, we detected it when observations were stacked\footnote{We stacked spectra by using addspec.py \url{https://github.com/JohannesBuchner/addspec.py}} (we did not include the observation with <0.5\,ks of exposure). Since then, all analyzed X-ray observations have yielded detections. 

\subsection{X-ray spectral modeling}
\label{subsec:xrayspec}
\renewcommand{\arraystretch}{1.3}

\begin{table*}[]
\caption{X-ray spectral fit results for \aarm}
    \centering

\begin{tabular}{lccc|cccccc}

\toprule
\toprule
      & \multicolumn{3}{c}{\textsc{powerlaw}} & \multicolumn{6}{c}{\textsc{diskbb+powerlaw}}      \\
\midrule
Epoch & $\Gamma$           & F$_{0.2-2}$      &  F$_{2-10}$    & kT & log(Norm)$_{\rm BB}$ & $\Gamma$ & \multicolumn{1}{l}{log(Norm)$_{\rm pow}$} &  F$_{0.2-2}$ &  F$_{2-10}$  \\
\midrule

CXO1   & $2.18^{+0.55}_{-0.59}$ & $1.47^{+0.92}_{-0.49}$ & $0.57^{+0.37}_{-0.23}$ & - & - & $2.17^{+0.61}_{-0.53}$ & $-5.54^{+0.20}_{-0.21}$ &$1.15^{+2.19}_{-0.58}$ & $0.57^{+0.36}_{-0.24}$ \\
EP1    & $1.20^{+0.36}_{-0.34}$ & $11.43^{+5.85}_{-3.44}$ & $8.86^{+5.78}_{-3.37}$ & - & - & $1.19^{+0.41}_{-0.37}$ & $-5.03^{+0.12}_{-0.14}$  &$3.50^{+2.28}_{-1.08}$ & $8.57^{+5.18}_{-3.67}$ \\
CXO2   & $2.10^{+0.30}_{-0.30}$ & $8.56^{+2.47}_{-1.59}$ & $3.57^{+1.13}_{-0.84}$ & - & - &  $2.08^{+0.30}_{-0.30}$ & $-4.80^{+0.12}_{-0.12}$  &$5.50^{+4.29}_{-1.74}$ & $3.55^{+1.12}_{-0.76}$ \\
EP2    & $2.31^{+0.42}_{-0.39}$ & $11.35^{+4.29}_{-2.79}$ & $3.06^{+2.28}_{-1.41}$ & - & - & $2.36^{+0.43}_{-0.44}$ & $-4.70^{+0.10}_{-0.12}$ &$10.03^{+6.63}_{-3.23}$ & $3.01^{+2.30}_{-1.36}$ \\
XMM1   & $1.98^{+0.11}_{-0.10}$ & $11.37^{+0.98}_{-0.90}$ & $4.80^{+0.73}_{-0.72}$ & $71^{+6}_{-7}$ & $2.96^{+0.30}_{-0.29}$ &  $1.76^{+0.14}_{-0.13}$ & $-4.81^{+0.04}_{-0.05}$  & $17.85^{+1.89}_{-1.77}$ & $5.61^{+0.96}_{-0.90}$ \\
EP3    & $1.55^{+0.47}_{-0.45}$ & $13.38^{+7.42}_{-4.40}$ & $8.29^{+6.99}_{-4.09}$ & - & - & $1.47^{+0.44}_{-0.42}$ & $-4.83^{+0.14}_{-0.14}$ &$5.08^{+2.94}_{-1.60}$ & $8.77^{+7.66}_{-4.10}$ \\
EP4    & $2.64^{+0.36}_{-0.36}$ & $15.90^{+5.45}_{-3.91}$ & $2.56^{+1.74}_{-1.08}$ & $48^{+12}_{-11}$ & $4.87^{+1.02}_{-0.87}$ & $2.61^{+0.43}_{-0.44}$ & $-4.64^{+0.09}_{-0.09}$ &$112.47^{+75.00}_{-45.04}$ & $2.45^{+1.78}_{-1.04}$ \\
EP5    & $3.32^{+0.28}_{-0.29}$ & $51.49^{+16.51}_{-11.65}$ & $2.19^{+1.07}_{-0.73}$ & $63^{+32}_{-37}$ & $3.32^{+1.36}_{-1.36}$ & $3.22^{+0.52}_{-0.84}$ & $-4.28^{+0.08}_{-0.14}$&$87.77^{+42.32}_{-20.98}$ & $2.52^{+3.38}_{-1.15}$ \\
EP6    & $3.41^{+0.27}_{-0.26}$ & $87.92^{+26.34}_{-18.44}$ & $3.08^{+1.30}_{-0.98}$ & $110^{+27}_{-21}$ & $2.55^{+0.58}_{-0.46}$ &   $2.11^{+1.50}_{-0.78}$ &  $-4.39^{+0.29}_{-0.26}$ &$91.74^{+29.85}_{-18.52}$ & $8.20^{+9.16}_{-5.79}$ \\
XMM2   & $3.77^{+0.40}_{-0.37}$ & $64.36^{+25.45}_{-17.60}$ & $1.03^{+0.73}_{-0.46}$ &  $87^{+8}_{-8}$ & $3.17^{+0.24}_{-0.23}$ & $2.11^{+0.65}_{-0.49}$ & $-4.6^{+0.14}_{-0.16}$ & $78.56^{+7.98}_{-7.29}$ & $6.29^{+4.24}_{-2.98}$ \\
CXO3   & $1.52^{+0.19}_{-0.19}$ & $17.80^{+2.44}_{-2.33}$ & $11.78^{+1.99}_{-1.99}$ & - & - & $1.49^{+0.20}_{-0.20}$ &  $-4.68^{+0.09}_{-0.09}$ &$7.21^{+66.10}_{-1.97}$ & $11.88^{+2.24}_{-2.01}$ \\
\bottomrule
\bottomrule
\label{tab:xrayfit}
\end{tabular}
\tablefoot{For each source, we report both the results of the single \textsc{powerlaw} spectral fit, as well as the \textsc{diskbb+powerlaw} model. In the majority of observations, the thermal parameters are unconstrained for the \textsc{diskbb+powerlaw} model. The temperature kT is expressed in eV. All fluxes are expressed in units of $10^{-14} \mathrm{erg/s/cm^{2}}$}
\end{table*}
\renewcommand{\arraystretch}{1.0}

\renewcommand{\arraystretch}{1.3}

\begin{table}[]
\caption{{X-ray spectral fit results for \aarm for the \textsc{simpl$\times$diskbb} model}.}
    \centering

\begin{tabular}{lcccc}

\toprule
\toprule
      
Epoch  & kT & log(Norm)$_{\rm BB}$ & $\Gamma$ & $f_{scatt}$  \\
\midrule

XMM1   & $69^{+7}_{-6}$ & $3.01^{+0.30}_{-0.28}$ & $1.77^{+0.14}_{-0.12}$ &  $0.050^{+0.013}_{-0.011}$  \\

EP4    & $55^{+16}_{-9}$ & $4.07^{+0.60}_{-0.84}$ & $2.71^{+0.37}_{-0.39}$ & $0.037^{+0.044}_{-0.015}$  \\
EP5    &  $72^{+23}_{-18}$ & $3.59^{+0.86}_{-0.76}$ & $2.70^{+0.53}_{-0.57}$ & $0.058^{+0.068}_{-0.028}$  \\
EP6    & $102^{+18}_{-28}$ & $2.84^{+0.76}_{-0.44}$ & $2.11^{+0.61}_{-0.75}$ &   $0.043^{+0.050}_{-0.018}$  \\
XMM2   &  $87^{+8}_{-8}$ & $3.19^{+0.24}_{-0.22}$ & $1.95^{+0.27}_{-0.19}$ & $0.023^{+0.019}_{-0.082}$  \\

\bottomrule
\bottomrule
\label{tab:xrayfit_phys}
\end{tabular}
\end{table}
\renewcommand{\arraystretch}{1.0}

We first attempted to fit the spectra in a homogeneous way, given the broad range of exposure times, total counts and bandwidths, testing both a powerlaw and multicolor blackbody models, each modified by Galactic absorption due to a line-of-sight column
density of N$_{\rm H}$ = 4.4 $\times10^{20} \mathrm{cm}^{-2}$, as estimated by the \citealp{Hi4pi}. In XSPEC language\footnote{all XSPEC models are available in SHERPA}, this is \textsc{const$\times$tbabs$\times$zashift$\times$pow}, or \textsc{const$\times$tbabs$\times$zashift$\times$diskbb}, where \textsc{zashift} is a convolution model which accounts for the redshift, which we set equal to z=0.0138, as reported in the DESI Data Release 1 (\citealp{beutler2026data}), and discussed in section \ref{sec:opt}. We fit all spectra between 0.5 and 10 keV, since all instruments are sensitive in this band.
A comparison of the Bayesian evidence $\mathrm{log(Z)}$ reveals that, across all observations, the powerlaw model is strongly preferred  ($\mathrm{\Delta log(Z) \geq 2}$). This is also supported by a visual inspection of the spectra and residuals, which are shown in Appendix \ref{app:xray-red}. We therefore report the spectral fits parameters for the powerlaw fit only in Table \ref{tab:xrayfit}.

The photon index of the single powerlaw model takes on values in the range $\Gamma \in[1.2-2.3]$ up to and including the February 11th observation EP3. This is significantly harder than most thermal TDEs (e.g., \citealp{Grotova2025b, Saxton2020}). The photon index then rises to the much more common range $\Gamma \in [2.6-3.8]$, until XMM2 on the 12th of March, while the latest Chandra observation CXO3, taken in April, reveals a spectral re-hardening. We refer to the different phases of the TDE as hard and soft states, respectively, depending on whether the photon index is below or above $\Gamma=3$. 

Even in the case of TDEs with hard X-ray spectra, there is often still evidence for soft X-ray emission arising from the innermost part of an accretion disk. Therefore, we extended our analysis of the XMM-Newton dataset down to 0.2 keV, as this instrument is uniquely sensitive at such low energies. Doing so revealed that a single powerlaw model was actually insufficient at modeling the XMM-Newton spectra between 0.2-10 keV, as the spectrum exhibited an excess of soft photons below 0.5 keV. We then modeled the spectra by adding a multicolor blackbody component to account for the soft emission (\textsc{const$\times$tbabs$\times$zashift$\times$(diskbb+pow)}), and found that this model reproduces the data well. The results of this modeling are shown in Table \ref{tab:xrayfit}. The spectra are shown in Section \ref{fis}, where we adopt a qualitatively analogous but more physically motivated model, which we describe in the next section.

As shown in Table \ref{tab:xrayfit}, which reports the results of this modeling, the blackbody parameters are virtually unconstrained in the majority of our observations, and in particular, for the Chandra observations. This is in agreement with the results of \citet{Guolo2024systematic} and \citet{chakramaster026}, which state that above a certain fraction of power-law contribution, it is impossible to recover TDE thermal parameters. However, adding a second model component, in addition to the power law, allows us to measure fluxes in a more physically motivated way. Therefore, in order to derive the accurate X-ray light curve shown in Fig. \ref{fig:lc}, we apply this model to all of our observations. We show the contour plots for the XMM1 and XMM2 fitting results in Fig. \ref{fig:contourbb}. The transition from the hard to the soft state is largely driven by changes in disk temperature and/or normalization, whereas the power-law parameters do not differ significantly. We do not show the results from EP4, 5, and 6, which are less constrained than those of XMM1 and 2, and are consistent with both these observations.

\begin{figure*}
    \centering
    \includegraphics[width=0.66\columnwidth]{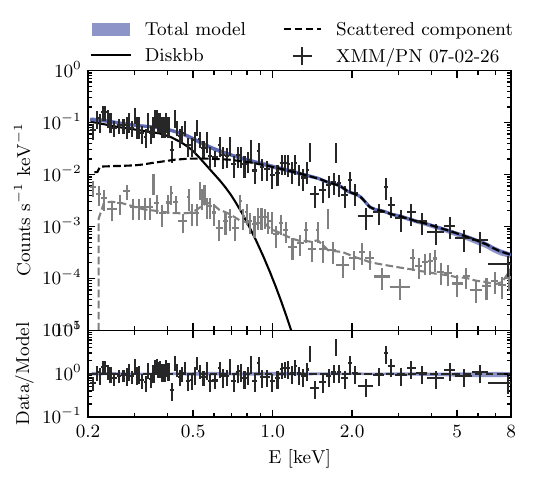}
    \includegraphics[width=0.66\columnwidth]{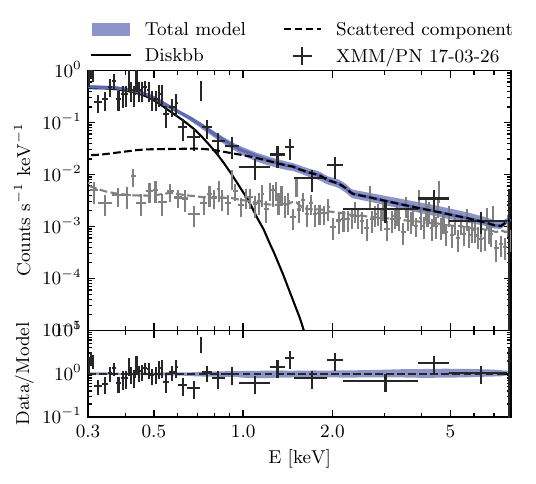}
    \includegraphics[width=0.66\columnwidth]{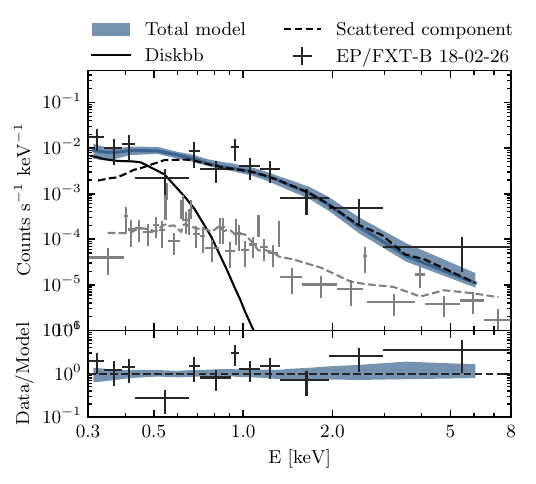} 
    \includegraphics[width=0.66\columnwidth]{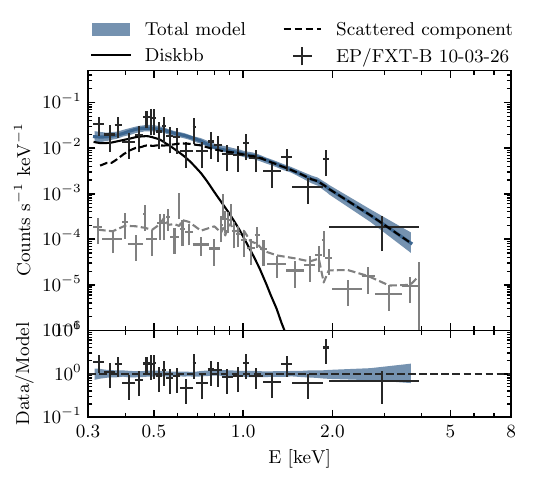}
    \includegraphics[width=0.66\columnwidth]{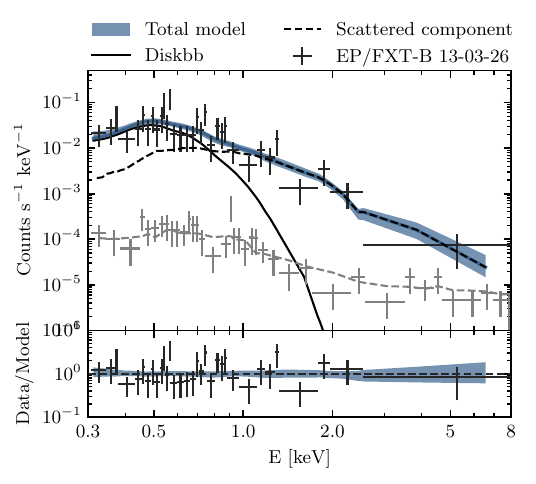}
     \includegraphics[width=0.66\columnwidth]{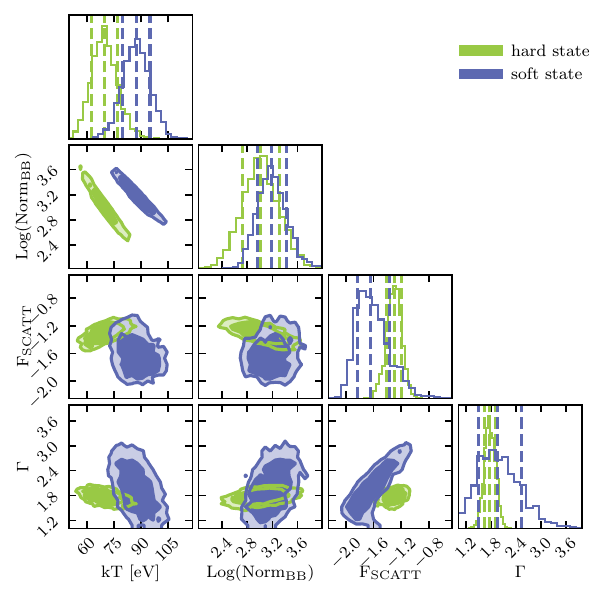}
    \caption{\textsc{simpl$\times$diskbb} model fit for XMM1, XMM2 and EP4,5, and 6, together with the XMM1 and XMM2 contour plots showing the 68th and 90th posterior percentiles.}
    \label{fig:complspec}
\end{figure*}

\subsection{Physically motivated modeling}
\label{fis}

\begin{figure*}
\centering
     \includegraphics[width=\linewidth]{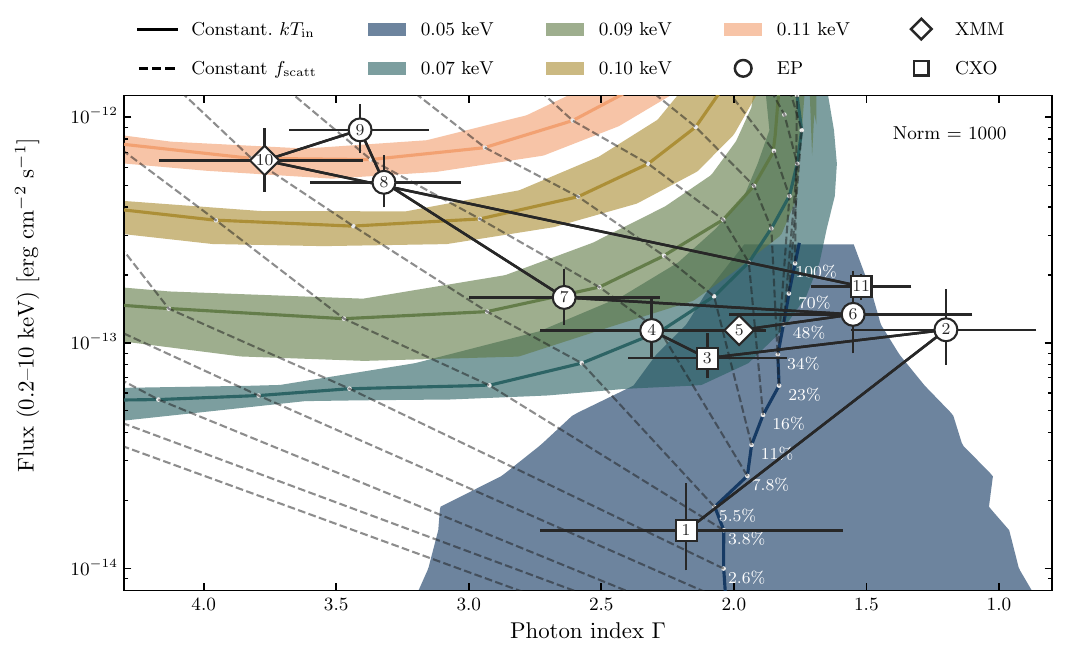}
    \caption{Flux versus $\Gamma$ evolution for \aarm. The black markers with white face color represent the different values explored by \aarm, as determined by the single power-law fit, numbered by epoch. The source appears to evolve counterclockwise, similarly to black hole XRB outbursts. The colored bands and the dashed lines show the loci that different combinations of physical parameters of the \textsc{simpl$\times$diskbb} model used in the simulations described in Section \ref{fis} would correspond to if the resulting spectra were fit with a single power-law. The colored curves and areas correspond each to a different input accretion disk temperature (see legend), and the dashed lines represent constant values of coronal scattering fractions (see the annotated percentage values). Refer to the text in section \ref{fis} for more details about the construction of this Figure. We also show the effect of changing disk normalization in Fig. \ref{fig:xrb_2}.}
    \label{fig:xrb}
\end{figure*} 

Next, we explored the possibility that the power-law component is physically linked to the disk blackbody. In particular, we modeled the hard component as the result of inverse-Compton upscattering of a fraction of seed thermal photons, similarly to a disk-corona system. At this scope, we replaced the powerlaw model with the \textsc{simpl} empirical Comptonization model, which upscatters a fraction $f_{scatt}$ of photons from a seed model (in this case \textsc{diskbb}) into a powerlaw spectrum. In this case, our model expression becomes \textsc{const$\times$tbabs$\times$zashift$\times$simpl$\times$diskbb} We apply this model to EP4, EP5, EP6, XMM1, and XMM2. The results of our fits can be seen in Table \ref{tab:xrayfit_phys}, and we show the spectra in Fig. \ref{fig:complspec}, together with the XMM1 and XMM2 contours. While, as in the case of the disjointed powerlaw and blackbody model, we find that the main driver of the hard and soft states is the disk, here it also appears, as shown by the contours in Fig. 2, that either the fraction of scattered photons or the Comptonizing powerlaw photon index has changed. 

As XMM1 and XMM2 are our best spectra, both in terms of bandwidth and net counts, we decide to explore whether the \textsc{simpl$\times$diskbb} could be the underlying model for all other observations. We therefore simulate X-ray spectra under different parameter settings for the \textsc{simpl$\times$diskbb} model and fit them with a single power law. At first order, comparing the photon indices and fluxes derived from the simulations with those of the actual observations will allow us to infer parameters for the \textsc{simpl$\times$diskbb} even when statistics would not allow it.  The purpose of the simulations is not parameter estimation but to visualize how changes in disk temperature and scattering fraction map onto the observed $\Gamma$–flux plane

The simulation setup is as follows: we simulate 500 source spectra for each parameter combination, informed by the results from XMM1 and XMM2, arbitrarily assuming 12\,ks of exposure and the EP/FXT-B response. We do so through the \textsc{fakeit} task of the \textsc{xspec} software package. We explore 20 values for $f_{SCATT}$ between 1\% and 100\%, uniformly sampled in logarithmic space, and 5 inner disk temperatures of kT=50, 75, 85, 100, and 110 eV. We fix the power-law photon index to 1.8 and run the simulation for two different \textsc{diskbb} normalizations: 1000 and 10000. We then fit a single power law to all spectra between 0.5 and 10 keV. For each parameter combination, we store the mean photon index and the 0.2-10 keV flux derived from the 500 spectra, along with their standard deviations.

We plot the results of the simulations in Fig. \ref{fig:xrb}. Different-colored lines each correspond to a constant temperature but an evolving scattering fraction, which are indicated as percentage values in the lowest-temperature simulations. The shaded areas represent the 1$\sigma$ uncertainties estimated from our Monte Carlo iterations. Conversely, dashed lines correspond to constant scattering fractions across different temperatures. In Fig. \ref{fig:xrb_2}, we show the effect of a varying disk normalization. As disk normalization increases, higher fluxes are observed for lower temperatures and lower scattering fractions, but the general trend remains the same.  

Finally, we overplot the photon indices and fluxes derived from the single powerlaw fit for all observations, using the same markers as in Fig. \ref{fig:lc} to distinguish between instruments. The temporal evolution is shown by the numbered markers.  \aarm initially moves upward in the plot, and then transitions from right to left as it reaches the highest flux values. Lastly, the last CXO3 reveals that the source has returned to the region characterized by hard spectra. 
We note that the XMM1 and XMM2 epochs lie in regions of the planes that are generally consistent with the values derived from detailed spectral fitting, though with slight offsets. In fact, the spectral fitting results of \aarm for the \textsc{simpl$\ast$diskbb} model show that the spectral transitions are due to the evolution of four different parameters: the disk temperature and normalization, as well as the scattering fraction and powerlaw photon index. Fig. \ref{fig:xrb} or Fig. \ref{fig:xrb_2} are only able to capture the role of two of these parameters at once. Moreover, for simplicity, we did not allow for the photon index $\Gamma$ to vary in our simulations. The figure is therefore powerful for studying the interplay of different components as the source evolves, but it is likely too simplistic to infer accurate parameter values.  

We note that this plot was constructed as a surrogate for the traditional hardness-intensity diagram (HID), commonly used for XRBs (e.g., \citealp{belloni2010}). XRB state transitions appear as Q-shaped loops in the HID, with total source counts on the y-axis and hardness ratio on the x-axis. As both hardness ratio and counts are instrument-dependent quantities, we decided to use the instrument-independent quantities of $\Gamma$ and flux, in order to qualitatively compare the evolution of \aarm to that of XRBs. The general evolution of \aarm over this plane can be described as moving from a low-hard state, to an intermediate-hard state, and lastly to a high-soft state, with a subsequent hardening and dimming. In terms of the assumed physical models, this implies an initial rise driven by the Comptonizing component, and a subsequent strengthening of the disk at the expense of the Comptonizing component. This is indeed reminiscent of the state transitions observed in X-ray binaries, as we further discuss in Section \ref{sec:xrb}.

\section{Optical analysis}
\subsection{Optical spectroscopy}
\label{sec:opt}

The host galaxy of \aarm was targeted by DESI in 2021, and the spectrum was released as part of the DESI Data Release 1, with a reported redshift of z=0.0138, which we adopted for this entire work (\citealp{beutler2026data}). We collected a co-added spectrum, representative of the pre-flare host galaxy. The spectrum, which is shown in Fig. \ref{fig:desi} appears to be red, and typical of quiescent galaxies. In order to derive information about the stellar kinematics, we analyze the spectrum using the Penalized Pixel-Fitting code \citealp{cappellari2004, cappellari2017}. First, we corrected the spectrum for Galactic foreground extinction assuming E(B-V)=0.0697 mag (\citealp{Schlafly2011}), and using  the Fitzpatrick extinction law extinction curve with R$_V$=3.1 (\citealp{fitzpatrick1999}). We applied this correction to all the spectra analyzed in this work. We used the flexible stellar population synthesis model (FSPS; \citealp{conroy2009, conroy2010}) as spectral templates, convolving them to match the DESI spectral resolution of R$\sim$2000-5500. We included gas templates and a 4th-order additive polynomial and performed the fit over the wavelength range 3599-9821 $\AA$. Following the prescriptions of \citealp{cappellari2012ppxf}, we derive the instrument-corrected stellar velocity dispersion $\sigma_*$=$86\pm2$ km/s. We estimate the SMBH black hole mass from the scaling relations of \citet{kormendy2013coevolution}, as  $\mathrm{log(M/M_\odot)}=6.87 \pm 0.31$. This is in agreement with the value derived by \citealp{simongini2026early}.

\begin{figure}
    \centering
     \includegraphics[width=1\linewidth]{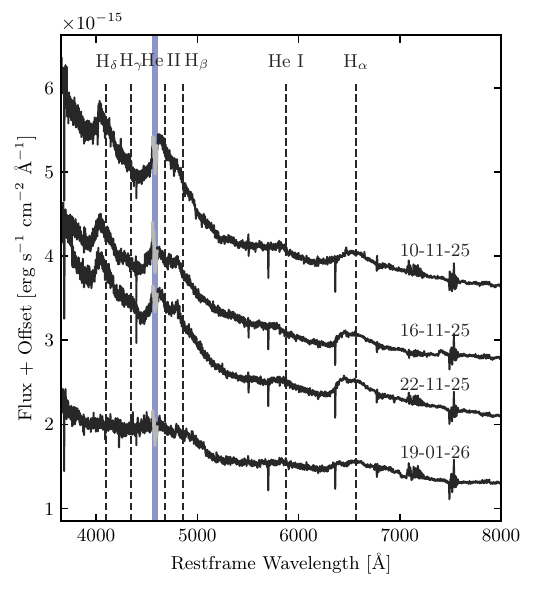}
    \caption{Host-subtracted follow-up optical spectra of \aarm. The blue vertical region marks the spectral range likely affected by spectral merging artefacts.}
    \label{fig:het}
\end{figure}

We also used the archival spectrum to subtract the host-galaxy contribution from the follow-up optical spectra obtained with HET, allowing us to study the transient-only spectral properties. In order to do so, we subtract the pPXF model of the DESI host-galaxy spectrum of \aarm from the follow-up spectra, interpolated to the spectral resolution of HET/LRS-2 ($R\sim1500$). For details about the HET/LRS-2 data collection and reduction, we refer to appendix \ref{app:opt}. We choose to use the model spectrum, in order to interpolate in a physically motivated fashion. For the host subtraction, we apply a multiplicative normalization factor chosen to remove the Ca H\&K absorption features from the transient spectra, as these are host-galaxy lines. The resulting host-subtracted spectra are shown in Fig. \ref{fig:het} (and the host-unsubtracted spectra in Fig. \ref{fig:het_2}).
As already reported in \citet{simongini2026early}, we detect broad H$\alpha$, He\,I, a broad complex including H$\beta$ and He\,II (and likely the Bowen line N\,III $\lambda$4641), and emission in correspondence of the H$\delta$ line. We note that we cannot conclusively discuss the presence of N\, III$\lambda$4641, as the line falls in a spectral range affected by artifacts from the merging of the two spectral arms of the LRS2 spectrograph. However, we also note that the significant detection of emission around the expected H$_\delta$ position, together with the non-detection of H$\gamma$, suggests that the emission around 4100 is actually dominated by the Bowen line NIII $\lambda$4100, rather than by H$\delta$ (e.g., \citealp{onori2019optical}). Therefore, we classify this TDE as a Bowen N\,III TDE. 

\begin{figure*}
    \centering
    \includegraphics[width=0.5\columnwidth]{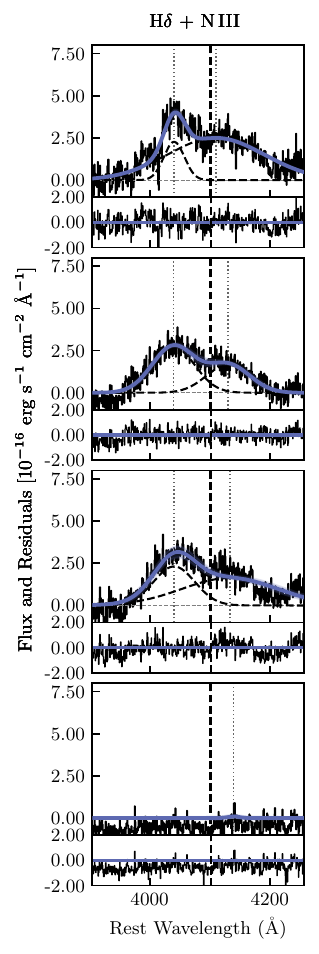}
    \includegraphics[width=0.5\columnwidth]{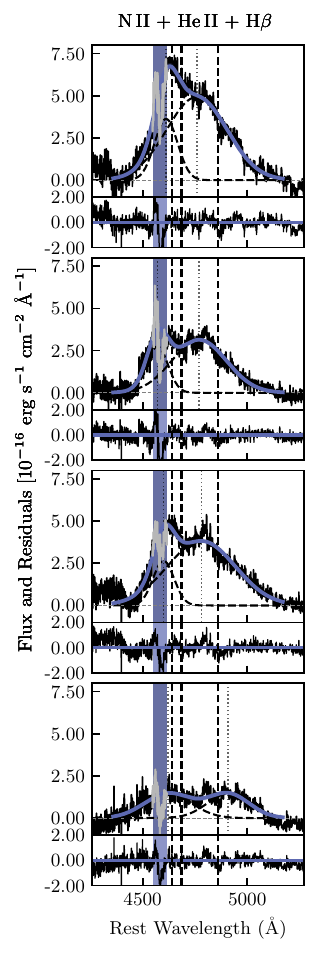}
    \includegraphics[width=0.5\columnwidth]{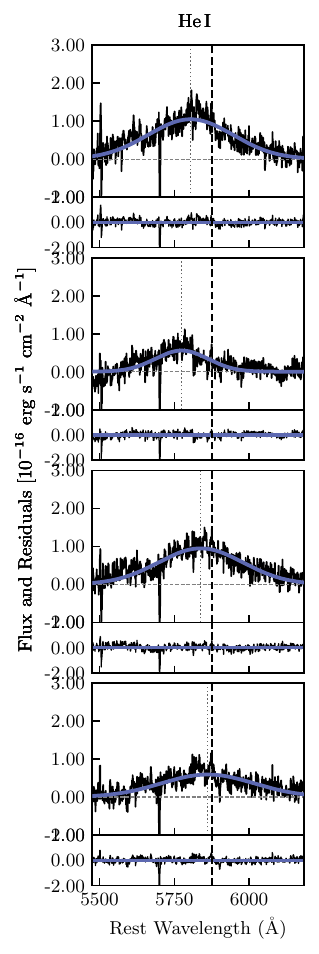}
    \includegraphics[width=0.5\columnwidth]{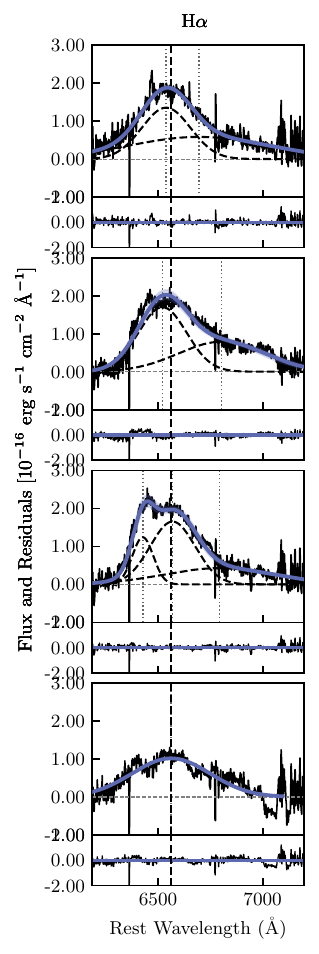}

    \caption{Fits of the optical spectral lines of \aarm. Each column shows a different line complex, and each row corresponds to a different epoch, increasing in descending order. The vertical dashed black lines indicate the expected restframe centroids for H$\delta$, N\,III, He\,II, H$\beta$,He\,I, and H$\alpha$, respectively, from left to right. The vertical gray dotted lines, instead, show the centroid of each fitted component. We show the spectral range likely affected by a spectral merging artifact in the shaded blue region.}
    \label{fig:optspec}
\end{figure*}

We then remove the transient continuum emission by fitting a 5th-order Legendre polynomial in line-free regions spanning 3780-3900, 4400-4450, 5300-5450, 6100-6200, 7700-7900, and 8250-8500 $\AA$. We then fit the residual emission lines with one to three Gaussians each using the lmfit Python package, following a procedure analogous to \citealp{baldini2025bowen}, adding up to three Gaussians for each complex as long as the Chi-square improves significantly. The results of our analysis are presented in Fig.~\ref{fig:optspec}, and are tabulated in Table \ref{tab:alllines}, where we report both the individual Gaussian measurements, as well as the total flux and FWHM for each line complex. Most emission lines, with the exception of He I, require two or three components in the first three epochs. The FWHMs of the emission lines are between 10000 and 30000 km/s, with the largest values being reached by the H$\beta$+He\,II(+N\,III) complex, where the emission is the result of the blending of two (and likely three) emission lines. The three pre-peak spectra show a blueshifted peak and a broad, smooth red shoulder, whereas the late-time spectrum shows more symmetric line profiles. This is consistent with the outflow profile evolution derived by \citealp{roth2016x} and observationally confirmed by the analysis of \citealp{charalampopoulos2022}.

\subsection{Host-galaxy spectral energy distribution modeling}

In order to understand the properties of the host galaxy of \aarm, we modeled the pre-flare spectral energy distribution (SED) from the MIR-to-UV. As the host galaxy is nearby, we collected the photometric information from the HECATEv2 \citealp{kyritsis2026hecatev2}, the second release of the Heraklion Extragalactic Catalog \citealp{kovlakas2021}, which is an all-sky value-added galaxy catalog including all the known galaxies within a volume of z$\lesssim$0.05 (D$\lesssim$200 Mpc). HECATEv2 provides expanded and homogenized optical and mid-infrared photometry from SDSS-DR17/NSA, PS1-DR2, and AllWISE surveys, accounting for extended galaxy sizes. In addition, HECATEv2 provides stellar population parameters (i.e., SFR, M$_{\star}$, metallicity) as well as activity classifications. As the current version of the HECATE catalog does not provide UV photometry, we adopted the NUV magnitude from the GALEX UV Unique Source Catalogs  (\citealp{bianchi2017revised}). We model the SED using photometry from GALEX, WISE, 2MASS, and SDSSDR17, through the SED-fitting
code Genuine Retrieval of AGN Host Stellar Population (GRAHSP, \citealp{buchner2024genuine}). 
We modeled the sed assuming a delayed star formation history, the \citet{bruzual203} stellar population syntheses models, nebular emission, all attenuated by dust and redshifted. The best-fit solution, shown in Fig. \ref{fig:SED}, corresponds to a stellar mass of $log(M_{*}/M_{\odot})= 10.6 \pm 0.1$, while the star formation rate is constrained only as an upper limit log(SFR$/(M_{\odot}/yr)$)<$-0.6$, consistent with a quiescent galaxy. By using the $M_{*}-M_{\bullet}$ scaling relations of \citealp{pucha2025}, we obtain a black hole mass from the derived $M_{*}$ of log($M_{\bullet}/M_{\odot}$)=$7.2\pm0.3$. This is consistent with the mass derived from $M_{\bullet}-\sigma_{*}$ relation, and from the mass reported by \citealp{simongini2026early} and derived by modeling the Optical-to-UV lightcuve of \aarm with MOSFiT (\cite{guillochon2018}), of log($M_{\bullet}/M_{\odot})=7.2\pm0.1 $.

\section{Discussion and conclusions}

\label{discussion}

 We have presented the X-ray and optical follow-up of the second closest TDE discovered to date, \aarm. While the optical spectroscopic and photometric evolution are representative of an average TDE, the X-ray properties are unlike classical events. For the first time, we detected a TDE transitioning from a low-hard to a high-soft X-ray state. Additionally, the X-ray luminosity of \aarm, measured by Chandra near the optical peak, makes this the faintest X-ray TDE ever detected close to the optical peak, and one of the TDEs with the highest optical to X-ray luminosity ratios. Before expanding on these points, however, we discuss whether the X-ray emission could, in fact, not be linked to the TDE.

\subsection{Contribution of an XRB population to the X-ray emission}

Before continuing the discussion, we explore the possibility that the hard X-ray emission at the location of \aarm does not arise from the TDE. In fact, galaxies with quiescent nuclei can still to be hard X-ray emitters, due to the presence of XRB populations. 
The number of X-ray binaries, and therefore the 2-10 keV X-ray output, correlates with the galaxy stellar mass (for low-mass XRBs, LMXRBs, e.g., \citealp{gilfanov2004}) and star formation rate (for high-mass XRBs, e.g., \citealp{bauer2002}). We therefore use Eq. 3 of \citealp{lehmer2010}, which relates the L$_{2-10}$ keV luminosity to the stellar mass $M_{*}$ and star formation rate SFR with a scatter of 0.34 dex, to estimate the predicted L$_{2-10\,keV}$ within the 1.5'' Chandra aperture. We first compute the stellar mass fraction within the Chandra aperture by adopting a Sersic profile with index $n=5.9$ and half-light radius of $R=7.0''$, as reported in the DESI Legacy Surveys data release 10 catalog (LS10, \citealp{dey2019}). The stellar mass in the Chandra aperture is $f\sim20\%$ of the total stellar mass of $log(M_{*})= 10.6 \pm 0.1$. The analysis returns an output of L$_{2-10\,keV}=7.1\pm0.3 \times 10^{38}$ erg/s, or F$_{2-10\,keV}=1.65\pm0.07$ erg/s/cm$^2$ in flux units. This is about a factor of 2.1-5.8 times lower than the observed 2-10 keV flux in CXO1, implying an XRB contribution of $<50\%$ (and less than a few percent in all other epochs), although the significance of this discrepancy is about 1.8$\sigma$. While statistically we cannot rule out that the hard X-ray component in CXO1 is due to a population of nuclear XRBs, the CXO1 excess with respect to the predicted L$_{2-10\,keV}$, as well as the detection of a consistent and steady rise of about a factor 20 of the hard X-ray component, with no significant spectral changes, supports the idea that the origin is related to the TDE.

We also tested the hypothesis that, instead of a population of XRBs, the nuclear X-ray emission is due to one or more ultra-luminous X-ray sources (ULXs, see e.g., \citealp{kaaret2017ultraluminous} for a review). The rate of ULXs in non-star-forming galaxies is N$_{ULX} = 6.3^{+1}_{-0.9}$ per 10$^{12} M_{\odot}$ (\citealp{kovlakas2020}). We calculate the Poissonian probability of finding 1 or more ULXs in the Chandra aperture, assuming that the distribution of ULXs traces the stellar mass distribution, as $(P\geq 1) = 1 - e^{-\ N_{ULX} \cdot M_* \cdot f}$, where $M_*$ is the total stellar mass. We derive a value of $(P\geq 1)=12\pm2\%$. While this is not negligible, it also does not appear to be the dominant scenario. The inclusion of the temporal and spectral evolution of the hard X-ray emission makes us effectively lean toward a TDE origin. A conclusive test of this hypothesis will require observing the nucleus of LEDA 3681212 once the TDE accretion flow has been completely dissipated. We proceed with the discussion under the assumption that the X-ray data we have collected are dominated at all phases by emission associated with the activity of \aarm, while highlighting the aforementioned caveats. 

\subsection{The exceptionally low early-time X-ray luminosity of \aarm }
\begin{figure}
    \centering

     \includegraphics[width=1\columnwidth]{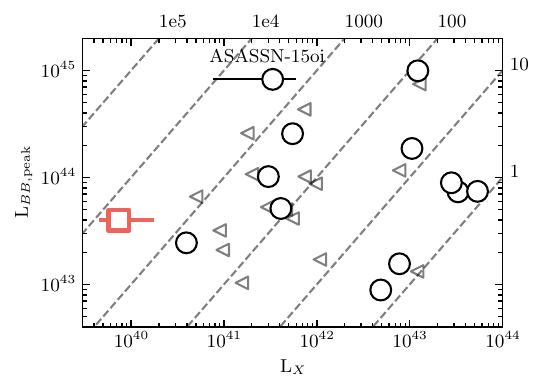}

    \caption{Peak optical bolometric blackbody luminosity versus early-time X-ray luminosity for a sample of TDEs. The red marker indicates the location of \aarm, for which we take the peak bolometric luminosity from \citealp{simongini2026early}. The X-ray luminosity is instead the one derived from the CXO1 observation in the 0.2-10 keV band. Black markers indicate the sample of optically bright and X-ray detected TDEs of \citealp{Guolo2024systematic}, and the gray triangles represent the sources from the same sample for which the X-ray luminosity could only be estimated as an upper limit. The luminosities and upper limits of \citealp{Guolo2024systematic} were derived in the 0.3-10 keV band.    }
    \label{fig:ratio}

\end{figure}

We plot the early time X-ray and peak optical luminosities for the optically selected TDEs from the sample of \citet{Guolo2024systematic} in Fig. \ref{fig:ratio}, together with \aarm. It can be easily seen that \aarm represents the faintest X-ray detection of an optically selected TDE close to optical peak to date, and that its peak optical-to-X-ray ratio of $L_{BB,\mathrm{peak}}/L_X=5.4 \pm 3.1 \times 10^3$ is the most extreme, together with that of ASASSN-15oi (\citealp{gezari2017}). It can be noted that the majority of X-ray upper limits for optically selected TDEs are at least one order of magnitude (and in some cases even three) higher than the early X-ray luminosity of \aarm. By referring to Fig. \ref{fig:lc}, one can appreciate that in the majority of cases, \aarm would not have been detected for $\sim$100 days after the optical peak.
The implications of such a detection strongly corroborate the point already presented in \citealp{Guolo2024systematic}: the historical dichotomy between X-ray bright and X-ray undetected TDEs is likely the result of selection effects (e.g., lack of deep, soft and long-lasting X-ray follow-up), which do not allow one to probe the highest end of the $L_{BB,\mathrm{peak}}/L_X$ distribution. Through the observations of \aarm, we show that this value can reach as high as $(5.4+3.1)\times10^{3}=8.5\times10^{3}$, if we include the uncertainty, and that weak accretion-driven X-ray emission might be more common than previously estimated.

\subsection{XRB-like state transitions}
\label{sec:xrb}

The most intriguing feature of \aarm, however, is its X-ray spectral evolution. While it is no longer surprising to observe TDEs showing a hard power-law component developing at later times in addition to the thermal emission from the accretion disk (e.g., \citealp{Wevers2021, hinkle2021discovery, Sfaradi2022, yao2022,cao2024, liu2023deciphering, berger2026,chakramaster026}), no thermal TDE to date has shown a transition from a power-law dominated state into a disk-dominated state, as its luminosity rises.

In section \ref{sec:xray}, we showed that \aarm climbs over the right side of the Flux vs. Photon index plane, and then moves leftward, as it further increases in flux. Lastly, the source appears to have dimmed and hardened again, in the last observation before sun-block. This evolution is qualitatively similar to the first half of a black hole X-ray binary outburst in the HID  (e.g., \citealp{belloni2010}). As we constructed the diagram as a surrogate for the HID in the case of multi-instrument observations, this suggests that evolution of \aarm somewhat resembles that of XRBs. This is further supported by our X-ray spectral-fitting results and simulations, which describe the evolution of \aarm as the interplay between the accretion disk and a Comptonizing corona, as in XRBs. 

As mentioned in the introduction, analogies between XRB state transitions and the variability of accretion flows onto SMBHs have been extensively researched, as they would constitute strong evidence that accretion is a scale-invariant process across several orders of magnitude in black hole mass. If we could confidently draw an analogy between black hole XRB outbursts and the evolution of \aarm, this would constitute one of the most compelling pieces of evidence supporting the scale invariance of accretion. Therefore, we must carefully discuss the caveats of this analogy. 
In particular, some works have noted that linearly scaling up the timescales of the XRB state transitions to the mass of the SMBH would imply much longer timescales than those observed for TDEs (e.g., \citealp{berger2026}). State transitions in XRBs occur on characteristic timescales of days to weeks. In the case of \aarm, this would correspond to thousands to tens of thousands of years, if scaled up from a $10 M_{\odot} $ to a $10^7 M_{\odot}$, as discussed in section \ref{sec:opt}. However, \citealp{goodwinmummery2026} noted that, while the linearity of the scaling should hold when comparing state transition timescales in XRBs and AGN, the picture is likely different when the SMBH is accreting in the aftermath of a TDE. This is because the viscous timescales, which are the timescales regulating the state transitions, can be proved to be independent of the black hole mass for both XRBs and TDEs. In fact, \citet{goodwinmummery2026} have shown that the expected timescales for state transitions in a TDE are a factor $\sim8$ longer than in the case of XRBs. The state transitions of \aarm, which occur on a timescale of $\sim100$ days, appear therefore to be fully consistent with this picture.

Additionally, XRB state transitions happen at specific Eddington ratios. In particular, the low-hard to high-soft state transitions occur at an Eddington ratio $\lambda_{\rm EDD}$ of about 2\% (\citealp{goodwinmummery2026}). In the case of \aarm, we can estimate the Bolometric luminosity and therefore the Eddington ratio by assuming that the X-ray-emitting accretion disk is the main power source of the TDE. As an order of magnitude estimation, we can therefore analytically derive the bolometric luminosity of \aarm from the \textsc{diskbb} model parameters using the following equation: 

\begin{equation}
\label{eg:1}
    L_{Bol} = 4\pi\sigma_{SB}R_{in}^2T_{in}^4 ,
\end{equation}

where $\sigma_{SB}$ is the Stefan-Boltzmann constant, $R_{in}$ is the inner disk radius, which is related to the \textsc{diskbb} normalization $N$, disk inclination angle $\theta$, and source distance $D$ as $N=(R_{in}/D_{10})^2 \cos\theta$, and $T_{in}$ is the inner disk temperature. If we calculate this for the XMM1 and XMM2 observations, as the representatives of the low-hard and high-soft X-ray states, we obtain bolometric luminositites in the ranges of $L_{Bol,XMM1} \sim [9 \times10^{41}-3\times10^{42}]$ and $L_{Bol,XMM2} \sim [3\times10^{42}-1\times10^{43}]$ (depending on the assumed inclination angle). For a $10^7 M_{\odot}$ BH, for which the Eddington luminosity is $1.26\times10^{45}$ erg/s, this would imply that the state transitions occur at $\sim0.3-1\%$ of Eddington, in agreement with expectations from the XRB analogy. This should be taken as an order of magnitude estimate, as we have neglected in Eq. \ref{eg:1} the correction factors discussed in \citealp{kubota1998evidence} and \citealp{shimura1995spectral}, as these have not been estimated for the case of a TDE. Moreover, while the origin of the optical emission in TDEs still remains uncertain (e.g., \citealp{mummery2025tidal}), it is possible that its contribution should be included in our estimations of $L_{Bol}$. A more robust estimation of $L_{BOL}$ for \aarm could be obtained with simultaneous broadband SED fitting (e.g., \citealp{guolo2025}). However, we do not expect our qualitative results to change significantly, as the inclusion of the optical luminosity estimated by the modeling \citet{simongini2026early} of $L_{diskBB}\sim1\times10^{43}$ at the time of the state transitions, would only confirm the expected transition threshold of $\lambda_{\rm EDD} \sim 2 \%$.

We note that $N$ can also provide an independent estimate of the SMBH mass. By assuming that $R_{in}$ corresponds to the SMBH's innermost stable circular orbit (ISCO), which can be expressed as $R_{ISCO} = f\times GM_{\bullet}/c^2$, where $G$ is the gravitational constant, and $c$ is the speed of light. The factor $f$ depends on the SMBH spin, and can have a value as low as 1 for prograde disks around maximally spinning Kerr SMBHs, of 6 for non rotating Schwarzschild SMBHs, and up to 9 for retrograde disks around maximally spinning Kerr SMBHS. Estimating the SMBH mass from the XMM1 and XMM2 normalizations, $N$ returns a range of $M_{\bullet}\in[9\times10^4,2\times10^6]$ due to our uncertainty on $\theta$ and the SMBH spin. While this can be up to two orders of magnitude lower than the masses obtained with MOSFiT and with host-galaxy scaling relations, the masses derived from the X-ray modeling can still be compatible with $\lambda_{\rm EDD}\sim$2\%, due to our uncertainty on $L_{Bol}$. We note that, in order for the diskbb mass estimate to be closer to the upper bound of the measured range, alleviating the discrepancy with the other mass estimates, both a high disk inclination angle $\theta$ and a spinning black hole are required.  

All in all, the state transitions of \aarm resemble those of XRBs, in which the spectral variations are driven by the interplay of the corona and the accretion disk. The timescales and regimes at which they occur are also in agreement with expectations from a unified accretion picture \citep{goodwinmummery2026}. We highlight that \aarm is the first TDE for which we are able to recover the rising branch of the typical XRB Q-shaped loop in the HID, and therefore represents one of the strongest SMBH analogs of stellar-mass accretors. We highlight that, if this is a common feature among TDEs, performing early-time XMM-Newton and Chandra observations of a nearby TDE classified before the optical peak, as in the case of \aarm, should allow one to observe such phenomenology again.  

We also note that it is possible that the hard X-ray component of \aarm is not due to a corona, but rather to the interaction of an outflow with the circumnuclear medium (e.g., \citealp{irwin2015,lei2016igr}). This has been explored for \aarm in \citealp{matsumoto2026faint}. The authors find that the X-ray evolution appears consistent with synchrotron emission from shocks between the expanding, unbound stellar debris and the circum-nuclear medium.  Such shocks are also expected to produce the radio detection reported by \citealp{christy2025}. We compare our observations with such predictions in Fig. \ref{fig:mp}, and find that significant differences between the X-ray evolution presented in this work and the predictions of \citealp{matsumoto2026faint}. However, it is possible that such a model could be expanded to reproduce the data presented in this work. Moreover, a proper investigation of this scenario will require the joint analysis of both radio and X-ray light curves, which is currently not feasible. 


\subsection{\aarm in the context of TDE models}

As shown in section \ref{sec:opt}, \aarm is a Bowen TDE.
Recent state-of-the-art simulations (e.g., \citealp{thomsen2026unified}) show that Bowen lines can be produced when X-ray emission encounters higher column densities of gas along the line of sight, and one would therefore expect some reprocessing of X-rays to occur at the same time. In the unified model proposed by \citealp{dai2018}, and extended with considerations on optical spectroscopic features in \citealp{charalampopoulos2022}, this reprocessing arises from an optically thick wind launched by the TDE disk, viewed at intermediate to large inclination angles.  At the same time, state-of-the-art TDE simulations (e.g., \citealp{lubonnerot, Martire2025, martire2026b}) have shown that stream collisions can also produce an optically thick envelope while the disk is still being assembled.
 
The exceptionally high $L_{BB,\mathrm{peak}}/L_X$ ratio observed in the early phase of \aarm is qualitatively consistent with both pictures. However, the observed X-ray state transitions imply that this extreme ratio is likely due to the disk being in an intrinsically fainter, low-Eddington state at early times, as our spectral modeling does not require absorption to reproduce the low-state spectra. At the same time, the hard X-ray spectra of TDE 2025aarm demonstrate that the source is capable of producing the ionization field necessary to generate Bowen lines, even in its low state. The presence of Bowen lines, therefore, indicates that a reprocessing layer exists, but it does not require that this layer dominate the observed optical emission. Interestingly, this seems to suggest that the bright optical flare in coincidence with the low-disk state might receive an additional power contribution from shocks, as suggested in \citealp{Martire2025,martire2026b}, though a direct comparison of our data with such a scenario is not possible because of the three orders of magnitude difference in black hole mass between \aarm and the mentioned simulations. 

\subsection{Conclusions}

The detection of low-hard to high-soft X-ray state transitions in \aarm, the faintest X-ray-detected TDE to date, represents a completely unprecedented phenomenology for TDEs. Such previously unprobed luminosities are opening the possibility that the X-ray behavior of \aarm could be a common feature of TDEs, previously unnoticed due to selection effects. This discovery constitutes positive evidence for the scale invariance of accretion onto stellar-mass and supermassive black holes. We stress that such a discovery would not have been possible without prompt, deep, and long-lived X-ray observations, and we advocate for more such observations of TDEs.       

\begin{acknowledgements}
PB would like to acknowledge Andrew Mummery and Paola Martire for the fruitful conversations. We are grateful to the XMM-Newton Project Scientist E. Kuulkers, to P. Slane and the Chandra X-ray Center team, for the DDT observations of \aarm. We are also grateful to the EP Science Center for executing the ToO observations of \aarm.
This work is based on data obtained with the Einstein Probe, a space mission supported by the
Strategic Priority Program on Space Science of the Chinese Academy of Sciences, in collaboration with the European Space Agency, the Max-Planck-Institute for extraterrestrial Physics (Germany), and the Centre National d’ ÅL Etudes Spatiales (France).
The Low Resolution Spectrograph 2 (LRS2) was developed and funded by the University of Texas at Austin McDonald Observatory and Department of Astronomy, and by Pennsylvania State University. We acknowledge the Texas Advanced Computing Center (TACC) at The University of Texas at Austin for providing high performance computing, visualization, and storage resources that have contributed to the results reported within this paper. We thank the Leibniz-Institut fur Astrophysik Potsdam (AIP) and the Institut fur Astrophysik Goettingen (IAG) for their contributions to the construction of the integral field units. Part of the funding for GROND (both hardware and personnel) was generously granted from the Leibniz-Prize to Prof. G. Hasinger (DFG grant HA 1850/28-1). 
P.C. acknowledges financial support from the Secretary of Universities and Research (Government of Catalonia) and by the Horizon 2020 Research and Innovation Programme of the European Union under the Marie Sk\l{}odowska-Curie and the Beatriu de Pin\'os 2024 BP 00125 programme, and from the Centro Superior de Investigaciones Cient\'ificas (CSIC) under the Spanish program Unidad de Excelencia María de Maeztu CEX2020-001058-M, financed by MCIN/AEI/10.13039/501100011033, and by the MaX-CSIC Excellence Award MaX4-SOMMA-ICE and acknowledges support via Research Council of Finland (grant 340613).
L.D. acknowledges support from the Hong Kong Research Grants Council (RGC GRF 17305124, 17305523) and the National Key R\&D Program of China (2025YFF0511100). 
TL acknowledges funding from the EU HORIZON-MSCA-2023-DN Project 101168906 ‘TALES: Time-domain Analysis to study the Life-cycle and Evolution of Super-massive black holes’.

\end{acknowledgements}
\bibliographystyle{aa} 
\bibliography{at2025aarm.bib}

\appendix

\section{Supplementary X-ray material}
\label{app:xray-red}
\subsection{X-ray spectra}

\begin{figure*}
 \includegraphics[width=0.5\linewidth]{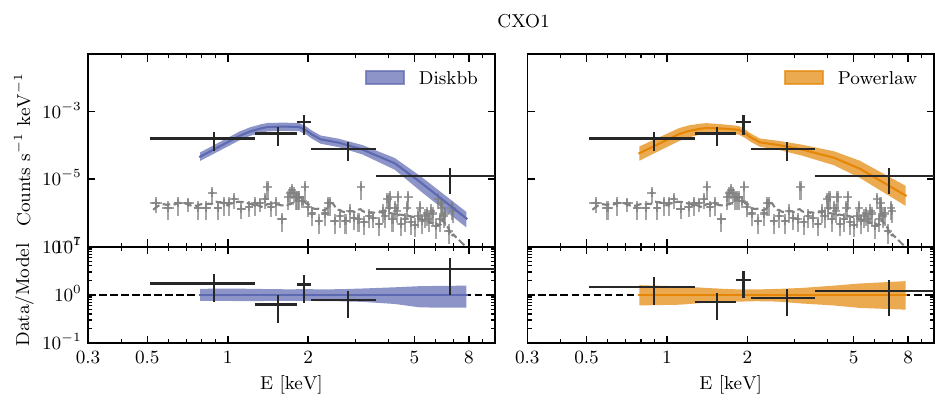}
     \includegraphics[width=0.5\linewidth]{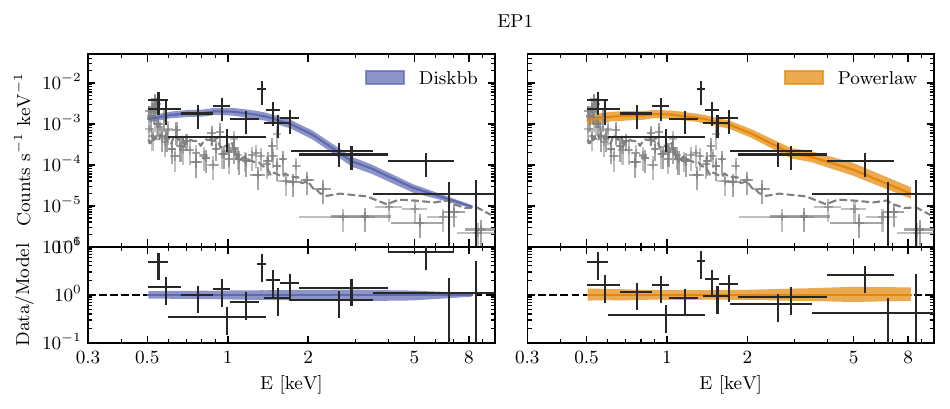}
     \includegraphics[width=0.5\linewidth]{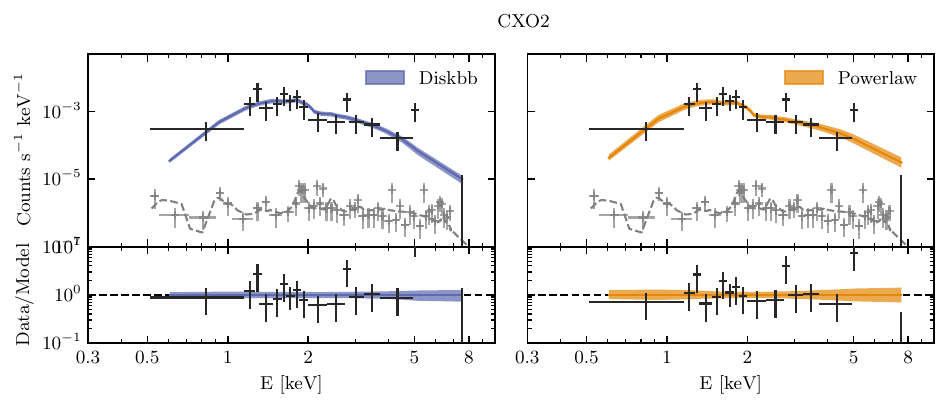}
     \includegraphics[width=0.5\linewidth]{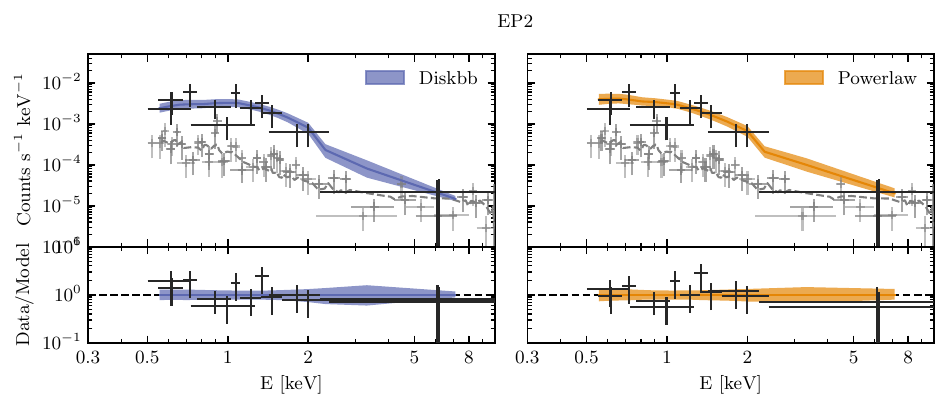}
     \includegraphics[width=0.5\linewidth]{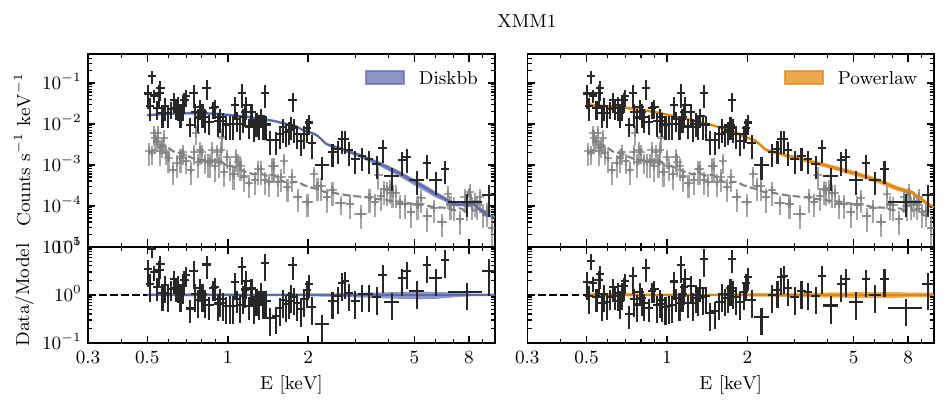}
      \includegraphics[width=0.5\linewidth]{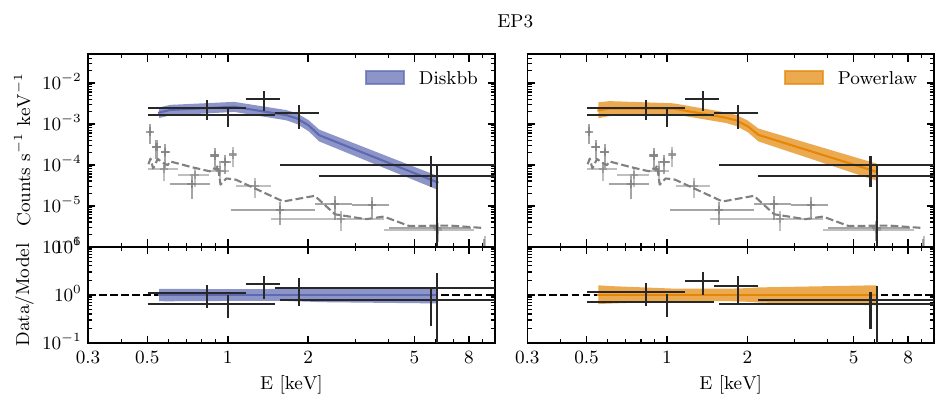}
     \includegraphics[width=0.5\linewidth]{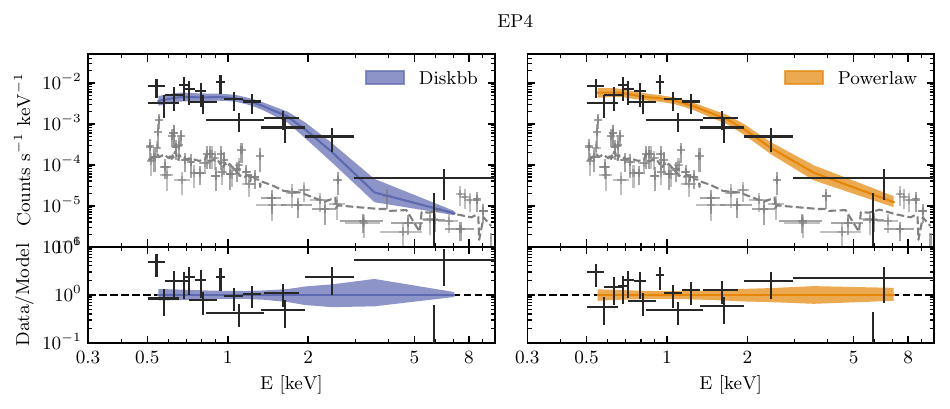}
     \includegraphics[width=0.5\linewidth]{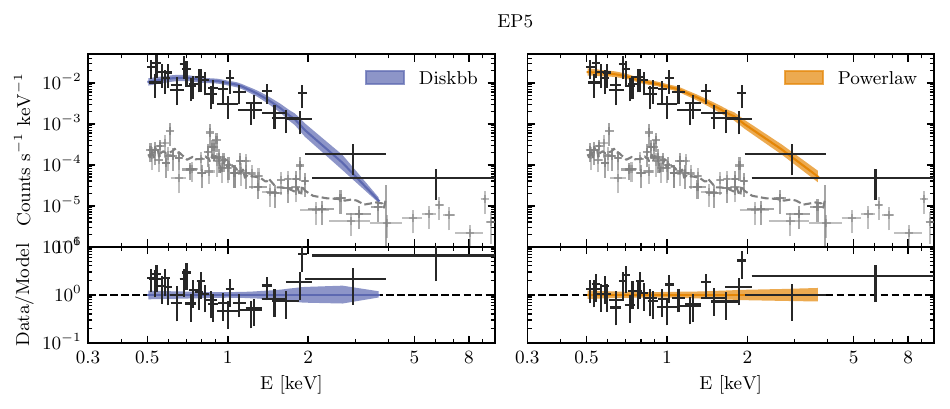}
     \includegraphics[width=0.5\linewidth]{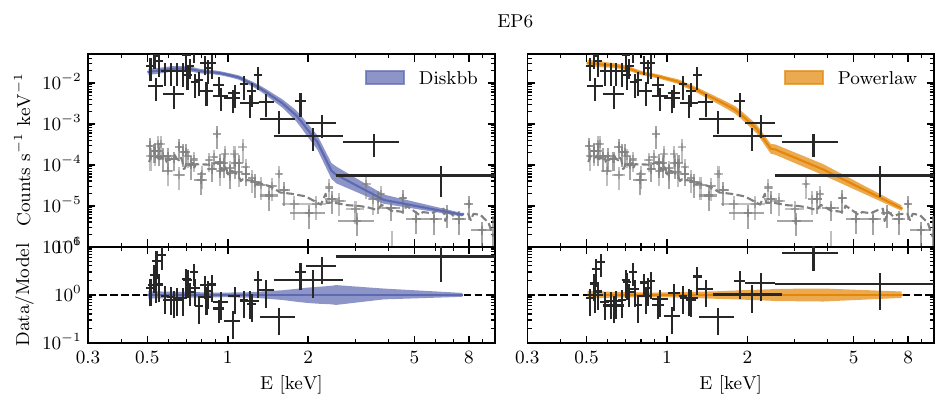}
     \includegraphics[width=0.5\linewidth]{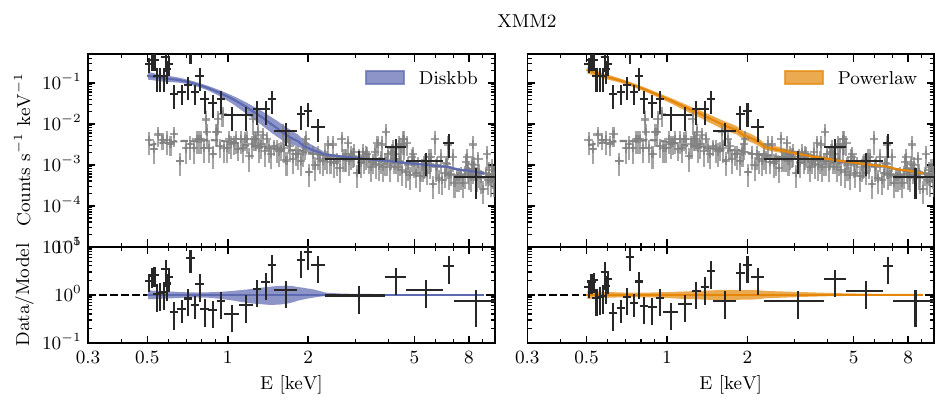}
     \includegraphics[width=0.5\linewidth]{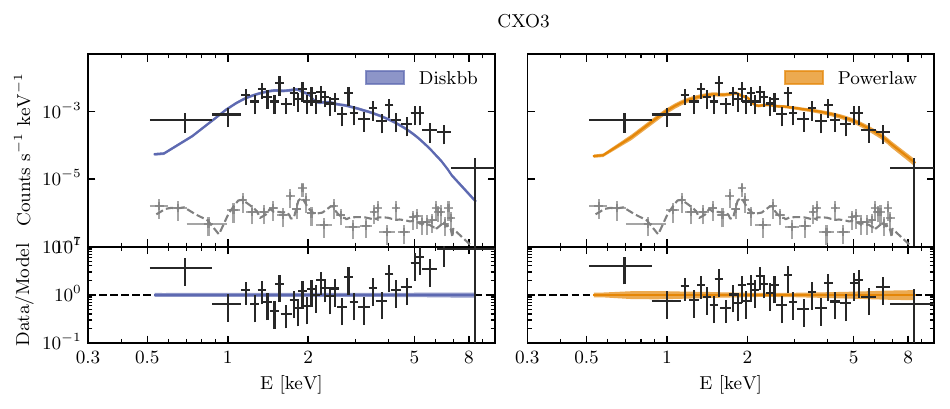}
     \caption{    All panels show, respectively, in black and gray, the source and background counts, rebinned for visualization purposes. the diskbb and powerlaw fits are respectively shown in the left and right panels. The gray line corresponds to the background model.}
    \label{fig:xrayspecall}
\end{figure*}

\begin{figure}
    \centering
 \includegraphics[width=1\columnwidth]{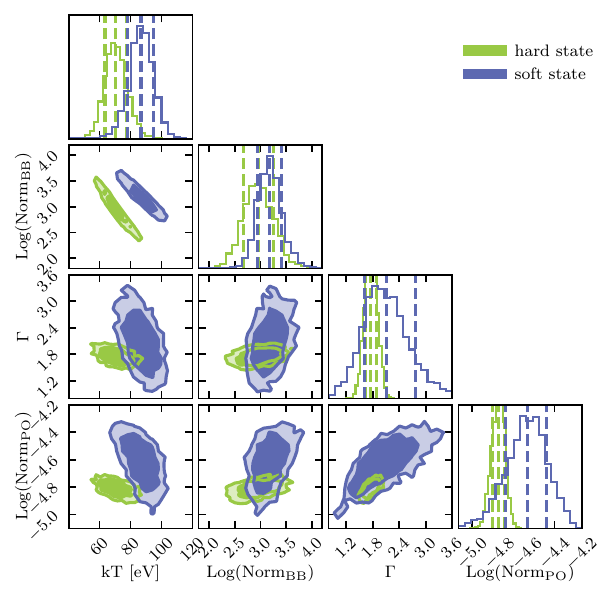}    
 \caption{XMM1 and XMM2 contour plots for the diskbb+powerlaw model. 
 }
 \label{fig:contourbb}
\end{figure}

\begin{figure}[h!]
    \centering
 \includegraphics[width=1\columnwidth]{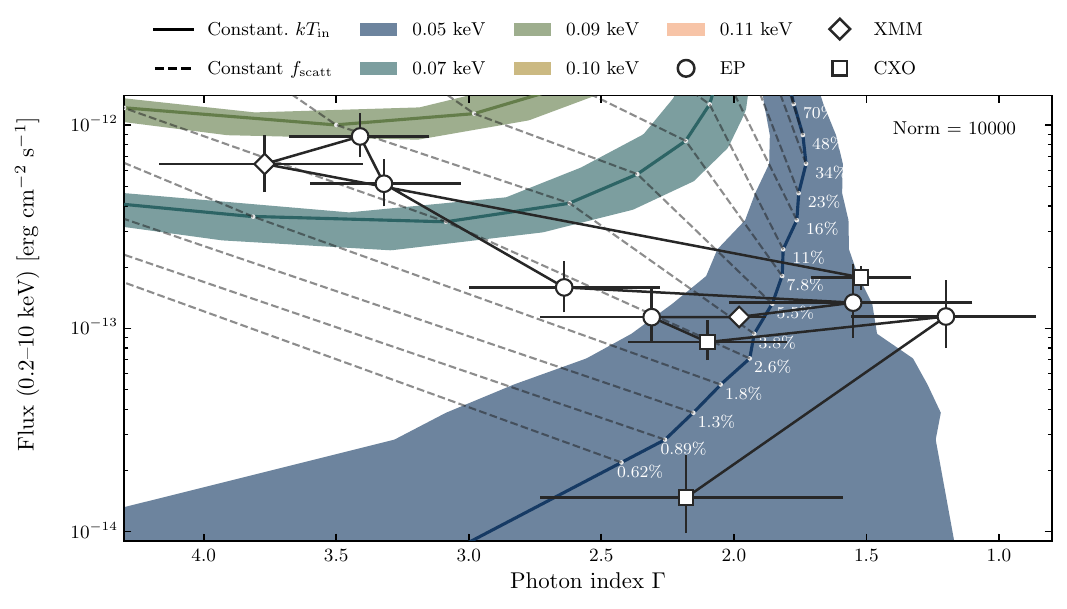}    
 \caption{Same as Fig. \ref{fig:xrb}, but with disk normalization 10000. The effect of a higher normalization is to produce the same photon index and flux as in the case with lower normalization, for lower values of disk temperature and scattering fraction.}
 \label{fig:xrb_2}
\end{figure}

In Fig. \ref{fig:xrayspecall}, we show all the X-ray spectra of \aarm, fit with a powerlaw and diskbb. As reported in the main text, in all cases, the powerlaw fit is highly preferred. In Fig. \ref{fig:contourbb}, we show the contour plots for the \textsc{diskbb+powerlaw} XMM1 and XMM2 fit. We also show in Fig. \ref{fig:xrb_2} the same plot shown in Fig. \ref{fig:xrb}, but with a higher input normalization.

\newpage
\break

\vfill
\FloatBarrier

\begin{figure}[h!]
    \centering
 \includegraphics[width=1\columnwidth]{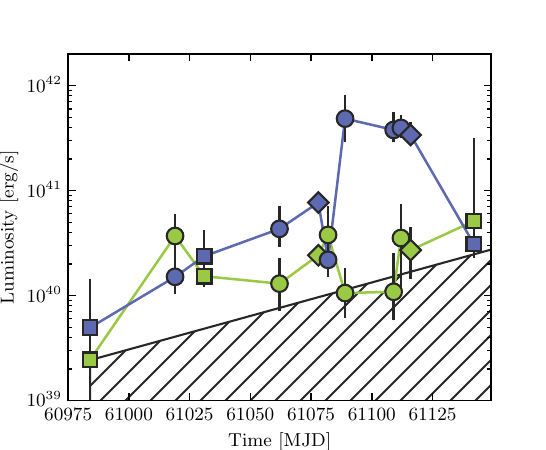}    
 \caption{The X-ray light curve of \aarm in luminosity space, compared to the predictions of \citealp{matsumoto2026faint}. The light curve is shown in the same way as in Fig. \ref{fig:lc}. All the predictions of \citealp{matsumoto2026faint} fall within the hatched area. Our observations show an excess of luminosity with respect to the predictions of \citealp{matsumoto2026faint} even when we only consider the 2-10 keV band, compared to the broadband used to predict the X-ray evolution in \citealp{matsumoto2026faint}.}
 \label{fig:mp}
\end{figure}

\FloatBarrier
\section{Optical analysis}
\label{app:opt}

\subsection{GROND data reduction and photometry extraction}

\aarm was observed with the Gamma-Ray burst Optical Near-infrared Detector (GROND; \citealt{Greiner:2008}) mounted at the MPG 2.2\,m telescope at ESO’s La Silla observatory in twelve epochs between 2026 January 17 and April 14th. Observations were performed simultaneously in the g$^{\prime}$, r$^{\prime}$, i$^{\prime}$ bands, J, H, and K$_{\rm s}$ bands. However, we here only report the results obtained in the g$^{\prime}$ (all epochs) and r$^{\prime}$ (two epochs) bands. The remaining r$^{\prime}$ data are corrupt due to a temporary technical issue with the detector. 

The data were processed with the standard IRAF-based GROND pipeline \citep{Kruehler:2015}. Image photometry was performed using AutoPhOT \citep{Brennan:2022} pipeline with the SFFT template subtraction package \citep{Hu2022SFFT}. Reference images obtained with Pan-STARRS in 2012 were used. For each image, the astrometric correction was derived using Astrometry.net \citep{Lang:2010}. An effective point-spread function (ePSF) model was constructed with the Photutils package \citep{Bradley:2024} from bright, isolated sources in each frame. Photometric zeropoints were calibrated using sequences of sources from the ATLAS All-Sky Survey Stellar Reference Catalog (REFCAT2, \citealt{Tonry:2018}.

\subsection{HET/LRS2 data collection and reduction}

HET/LRS2 is an instrument composed of 2 spectrographs, LRS2-B and LRS2-A, covering respectively $3700-7000 \AA$ and $6500-10500 \AA$. The spectrographs are integral field units, each composed of a bundle of 0.6'' fibers covering 12''$\times$6''. The observations are carried out by an astronomer on site and are reduced at the Texas Advanced Computing Center, using the \textsc{panacea} pipeline (\citealp{zeimann2026panacea}). Full details on the general
data-reduction procedures are provided in \citealp{chonis2016lrs2}.
We further reduce the flux calibrated outputs of the \textsc{panacea} pipeline through the \textsc{LRS2Multi} package\footnote{\url{https://github.com/grzeimann/LRS2Multi}}, which allows to refine sky subtraction and source extraction. We choose to extract source 1D spectra for each epoch using a 0.75'' radius circular aperture, in order to match the diameter of DESI fibers. The sky is estimated from fibers with radii larger than 4'' from the source and is subtracted from the source spectra. We match the normalizations of the LRS2-R and LRS2-B spectra over the range 6800$\AA$ to 7000$\AA$. 

\subsection{complementary figures}
\begin{figure}[h!]
    \centering
     \includegraphics[width=1\linewidth]{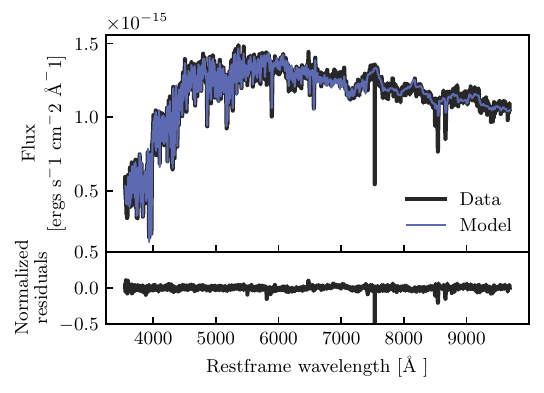}
    \caption{pPXF spectral fitting results (blue line) for the DESI spectrum (black line) of \aarm, together with residuals.}
    \label{fig:desi}
\end{figure}

\begin{figure}[h!]
    \centering
     \includegraphics[width=1\linewidth]{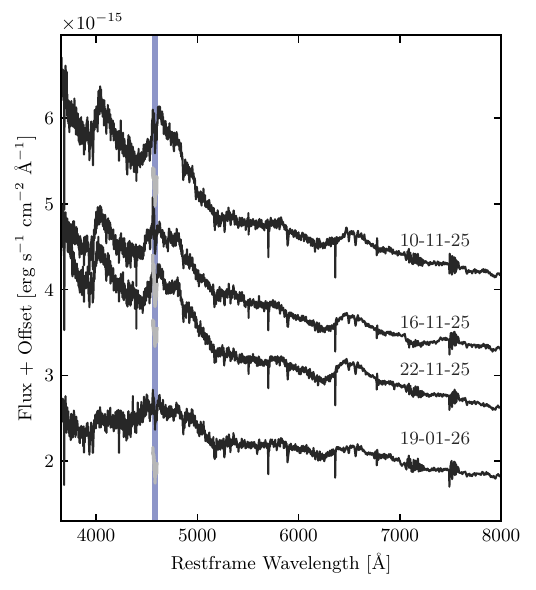}
    \caption{The HET spectra prior to host subtraction.}
    \label{fig:het_2}
\end{figure}
\begin{figure}
    \centering

     \includegraphics[width=1\linewidth]{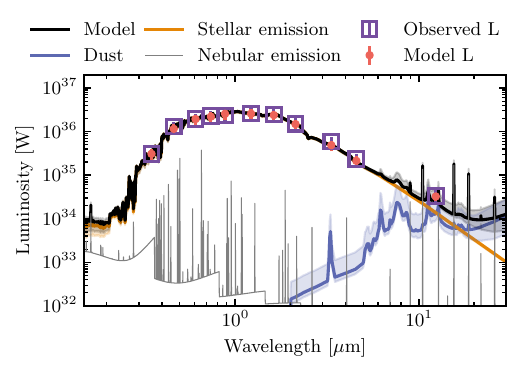}

    \caption{Observed and modeled SED of \aarm. The purple squares represent the luminosities derived from the GALEX, SDSSDR17, WISE, and 2MASS observed
fluxes, while the red dots represent the model-predicted luminosities. The orange, gray, and blue lines represent the best-fit individual model components, which are respectively stellar, nebular, and dust emission. The black line is the total best-fit model.}
    \label{fig:SED}

\end{figure}

\begin{table*}
\centering
\caption{Line measurements for each of the 4 follow-up spectra.}
\label{tab:alllines}
\begin{tabular}{clccccc}
\toprule
\toprule
ID & Parameters & Comp. & H$\delta$(+NIII) & H$\beta$+He\,II(+N\,III) & He\,\textsc{i} & H$\alpha$ \\
\hline
1  & Centroid (\AA)          & 1 &  $ 4110.1 \pm 3$ & $ 4758.9 \pm 2.8 $ & $ 5804.1 \pm 3 $ & $ 6538.4 \pm 2.3$ \\
   &                         & 2 & $ 4040.7 \pm 1.1$ & $ 4611.8 \pm 1.2 $ & - & $ 6694.5 \pm 13.5$  \\
   &                         & 3 & - & - & - & - \\
\cline{2-7}
   & FWHM (km s$^{-1}$)      & 1 &   $13820 \pm 390$  & $21460 \pm 240$ & $17127 \pm 371$ & $12612 \pm 331$ \\
   &                         & 2 & - & $2996 \pm 234$ & - & $35473.\pm 1207$\\
   &                         & 3 & - & - & - & - \\
\cline{2-7}
   & Flux                    & 1 & $5.1 \pm 0.19 \times 10^{-14}$ & $17.87 \pm 0.27 \times 10^{-14}$ & $3.75 \pm 0.1 \times 10^{-14}$ & $3.96 \pm 0.15 \times 10^{-14}$ \\
   &                         & 2 & $0.97 \pm 0.09 \times 10^{-14}$ & $5 \pm 0.21 \times 10^{-14}$  & - & $4.51 \pm 0.25 \times 10^{-14}$ \\
   &                         & 3 & - & - & - & - \\
\cline{2-7}
   & Total FWHM (km s$^{-1}$) &  & $11000 \pm 522$ & $22031 \pm 330$ & $17127 \pm 371$ & $15826 \pm 450$ \\
   & Total Flux               &  & $6.07 \pm 0.2 \times 10^{-14}$  & $22.87 \pm 0.31 \times 10^{-14}$  &  $3.75 \pm 0.1 \times 10^{-14}$ & $7.56 \pm 0.21 \times 10^{-14}$ \\
\hline

2  & Centroid (\AA)          & 1 & $ 4130.6 \pm 3.5$ & $ 4769.8 \pm 3.3$ & $ 5774.4 \pm 3.7$ & $ 6434.1\pm 1.3$\\
   &                         & 2 & $ 4039.2 \pm 2.6$ & $ 4570.4 \pm 1.2$ & - & $ 6577.2.1 \pm 8$ \\
   &                         & 3 & - & - & - & $ 6882.2 \pm 18$ \\
\cline{2-7}
   & FWHM (km s$^{-1}$)      & 1 & $6145 \pm 582$ & $19306 \pm 434$ & $10813 \pm 458 $ &  $5486 \pm 320$ \\
   &                         & 2 & $6153 \pm 351$ & $7743 \pm 270$ & - & $11998 \pm 575$ \\
   &                         & 3 & - & - & - & $17838 \pm 1202$ \\
\cline{2-7}
   & Flux                    & 1 & $1.51 \pm 0.16 \times 10^{-14}$ & $10.28 \pm 0.25 \times 10^{-14}$ & $1.23 \pm 0.07 \times 10^{-14}$ & $1.26 \pm 0.12 \times 10^{-14}$ \\
   &                         & 2 & $2.43 \pm 0.15 \times 10^{-14}$& $3.45 \pm 0.17 \times 10^{-14}$ & - & $4.73 \pm 0.31 \times 10^{-14}$ \\
   &                         & 3 & - & - & - &  $2.88 \pm 0.22 \times 10^{-14}$ \\
\cline{2-7}
   & Total FWHM (km s$^{-1}$) &  & $ 11395 \pm 528$ & $25368 \pm 405$ & $10813 \pm 458 $ & $17748  \pm 1023$ \\
   & Total Flux               &  & $3.94 \pm 0.18 \times 10^{-14}$  & $13.62 \pm 0.28 \times 10^{-14}$ &  $1.23 \pm 0.07 \times 10^{-14}$ & $8.87 \pm 0.33 \times 10^{-14}$ \\
\hline

3  & Centroid (\AA)          & 1 &  $  4133.3 \pm 12.3$  & $  4599.6 \pm 1.3$ & $ 5838.0 \pm 2.8$ & $  6426.7 \pm 1.8$ \\
   &                         & 2 &  $  4040.7 \pm 1.7$ & $ 4780.2 \pm 2.9$ & - & $ 656925 \pm 6.8$\\
   &                         & 3 & - & - & - & $ 6792 .5\pm 8.8$ \\
\cline{2-7}
   & FWHM (km s$^{-1}$)      & 1 & $13543 \pm 1281$& $7688 \pm 272$ &  $17178 \pm 338$ & $5492 \pm 267$ \\
   &                         & 2 & $6010 \pm 541$ & $23983 \pm 309$ & - & $11528 \pm 674$ \\
   &                         & 3 & - & - & - & $ 27375 \pm 1294$ \\
\cline{2-7}
   & Flux                    & 1 & $3.23 \pm 0.31 \times 10^{-14}$ & $3.23 \pm 0.16 \times 10^{-14}$ & $3.38 \pm 0.09 \times 10^{-14}$ & $1.55 \pm 0.16 \times 10^{-14}$ \\
   &                         & 2 & $1.97 \pm 0.3 \times 10^{-14}$ & $1.55 \pm 0.24 \times 10^{-14}$ & - & $4.44 \pm 0.31 \times 10^{-14}$ \\
   &                         & 3 & - & - & - & $2.81 \pm 0.25 \times 10^{-14}$ \\
\cline{2-7}
   & Total FWHM (km s$^{-1}$) &  & $11318 \pm 2015 $ & $26065 \pm 394$ & $17178 \pm 338$ & $15805 \pm 670$ \\
   & Total Relative Flux      &  & $5.2 \pm 0.39 \times 10^{-14}$ & $4.78 \pm 0.26 \times 10^{-14}$ &  $3.38 \pm 0.09 \times 10^{-14}$ & $8.79 \pm 0.32 \times 10^{-14}$ \\
\hline

4  & Centroid (\AA)          & 1 & - & $ 4621 \pm 6 $  & $ 5861.6 \pm 5 $ & $ 6561.8\pm 2.86$\\
   &                         & 2 & - & $ 4908 \pm 5.3 $ & - & - \\
   &                         & 3 & - & - & - & - \\
\cline{2-7}
   & FWHM (km s$^{-1}$)      & 1 & - & $ 17077 \pm 786$ &  $18700 \pm 615$ & $20096 \pm 313$ \\
   &                         & 2 & - & $14368 \pm 615$ & - & - \\
   &                         & 3 & - & - & - & - \\
\cline{2-7}
   & Flux                    & 1 & - & $4.12 \pm 0.21 \times 10^{-14}$ & $2.31 \pm 0.09 \times 10^{-14}$ & $4.77 \pm 0.1 \times 10^{-14}$ \\
   &                         & 2 & - &  $3.62 \pm 0.17 \times 10^{-14}$ & - & - \\
   &                         & 3 & - & - & - & - \\
\cline{2-7}
   & Total FWHM (km s$^{-1}$) &  & - & $32455\pm 694$  &  $18700 \pm 615$ & $20096 \pm 313$ \\
   & Total Flux               &  & - & $7.74 \pm 0.23 \times 10^{-14}$ & $2.31 \pm 0.09 \times 10^{-14}$ & $4.77 \pm 0.1 \times 10^{-14}$ \\
\bottomrule
\bottomrule
\end{tabular}

\end{table*}

\end{document}